\documentstyle[preprint,eqsecnum,prd,aps]{revtex}
\input{epsf}
\begin{document} 
% states and fields
\renewcommand{\thefootnote}{\fnsymbol{footnote}}
\setcounter{equation}{0}
\newcommand{\beq}{\begin{equation}}
\newcommand{\eeq}{\end{equation}}
\newcommand{\beqa}{\begin{eqnarray}}
\newcommand{\eeqa}{\end{eqnarray}}
\pagestyle{plain}
%---------------------------------------------------------------------------
%------------------NEW ADDITIONS TO EXISTING ARTICLE.STY------------------
\catcode`\@=11
\long\def\@makefntext#1{
\protect\noindent \hbox to 3.2pt {\hskip-.9pt  
$^{{\ninerm\@thefnmark}}$\hfil}#1\hfill}		%CAN BE USED 

\def\@makefnmark{\hbox to 0pt{$^{\@thefnmark}$\hss}}  %ORIGINAL 
	
\def\ps@myheadings{\let\@mkboth\@gobbletwo
\def\@oddhead{\hbox{}
\rightmark\hfil\ninerm\thepage}   
\def\@oddfoot{}\def\@evenhead{\ninerm\thepage\hfil
\leftmark\hbox{}}\def\@evenfoot{}
\def\sectionmark##1{}\def\subsectionmark##1{}}

%--------------------START OF PROCSLA.STY---------------------------------
% For symbolic footnotes indices in title/author preamble
\setcounter{footnote}{0}
\renewcommand{\thefootnote}{\fnsymbol{footnote}}
					         
%NEW MACRO TO HANDLE APPENDICES
\newcounter{appendixc}
\newcounter{subappendixc}[appendixc]
\newcounter{subsubappendixc}[subappendixc]
\renewcommand{\thesubappendixc}{\Alph{appendixc}.\arabic{subappendixc}}
\renewcommand{\thesubsubappendixc}
	{\Alph{appendixc}.\arabic{subappendixc}.\arabic{subsubappendixc}}

\renewcommand{\appendix}[1] {\vspace*{0.6cm}
        \refstepcounter{appendixc}
        \setcounter{figure}{0}
        \setcounter{table}{0}
        \setcounter{equation}{0}
        \renewcommand{\thefigure}{\Alph{appendixc}.\arabic{figure}}
        \renewcommand{\thetable}{\Alph{appendixc}.\arabic{table}}
        \renewcommand{\theappendixc}{\Alph{appendixc}}
        \renewcommand{\theequation}{\Alph{appendixc}.\arabic{equation}}
%       \noindent{\bf Appendix \theappendixc. #1}\par\vspace*{0.4cm}}
        \noindent{\bf Appendix \theappendixc #1}\par\vspace*{0.4cm}}
\newcommand{\subappendix}[1] {\vspace*{0.6cm}
        \refstepcounter{subappendixc}
        \noindent{\bf Appendix \thesubappendixc. #1}\par\vspace*{0.4cm}}
\newcommand{\subsubappendix}[1] {\vspace*{0.6cm}
        \refstepcounter{subsubappendixc}
        \noindent{\it Appendix \thesubsubappendixc. #1}
	\par\vspace*{0.4cm}}

%---------------------------------------------------------------------------
%FOLLOWING THREE COMMANDS ARE FOR `LIST' COMMAND.
\topsep=0in\parsep=0in\itemsep=0in
\parindent=1.5pc

%---------------------------------------------------------------------------
%FIGURE CAPTION
\newcommand{\fcaption}[1]{
        \refstepcounter{figure}
        \setbox\@tempboxa = \hbox{\footnotesize Fig.~\thefigure. #1}
        \ifdim \wd\@tempboxa > 6in
           {\begin{center}
        \parbox{6in}{\footnotesize\baselineskip=12pt Fig.~\thefigure. #1}
            \end{center}}
        \else
             {\begin{center}
             {\footnotesize Fig.~\thefigure. #1}
              \end{center}}
        \fi}

%TABLE CAPTION
\newcommand{\tcaption}[1]{
        \refstepcounter{table}
        \setbox\@tempboxa = \hbox{\footnotesize Table~\thetable. #1}
        \ifdim \wd\@tempboxa > 6in
           {\begin{center}
        \parbox{6in}{\footnotesize\baselineskip=12pt Table~\thetable. #1}
            \end{center}}
        \else
             {\begin{center}
             {\footnotesize Table~\thetable. #1}
              \end{center}}
        \fi}

%%%%Start of Text%%%%%%%%%%%%%%%%%%%%%%%%%%%%%%%%%%%%%%%%%%%%%%%%%%%%%%%%%%%%
%\rightline{
\preprint{
\vbox{
\halign{&##\hfil\cr
	& ANL-HEP-PR-96-33\cr
        & MC-TH-96-16\cr}}
}
%}
\title{New Parton Distribution Functions for the Photon}

\author{L. E. Gordon$^a$ and J. K. Storrow$^b$}
\address{
$^a$ High Energy Physics Division, Argonne National Laboratory,
	Argonne, IL 60439, USA \\ $^b$ 
Dept. of Theoretical Physics, University of Manchester, 
Manchester M13 9PL, England}

\maketitle
\begin{abstract} 
We present new improved parton distributions for the photon. 
We fit {\bf all} available data on the photon structure function, 
$F^{\gamma}_{2}(x,Q^2)$, with $Q^2\ge 3$ GeV$^2$, in order to 
determine the quark distributions. We also pay particular attention  
to the gluon distribution in the photon, $g^{\gamma}(x,Q^2)$, 
which has been poorly constrained in earlier 
analyses which only include structure function data. We use large $p_T$ 
jet production in $\gamma \gamma$ collisions from TRISTAN to constrain 
$g^\gamma $. We also see what information can be gleaned from $\gamma p$ 
collisions at HERA on $g^{\gamma}$ and on the quark distributions at large 
$x$, where there are no structure function data. We review future 
prospects of elucidating the parton distributions of the photon.   
\end{abstract}
\vspace{0.2in}

\narrowtext

\section{Introduction}

Some time ago, we presented some parton distribution functions (pdfs)
for the photon\cite{GS}. These distributions, which we will refer to 
here as the GS distributions, were the first available in both 
leading order(LO) and next-to-leading order(NLO) QCD. Since then they have 
been used as input to perturbative QCD (PQCD) comparisons to a variety of 
types of data with a fair degree of success. Applications 
in $\gamma \gamma $ collisions at various $e^+e^-$colliders are not
restricted to the analysis of $F^{\gamma}_{2}(x,Q^2)$
data\cite{DELPHI1}; they also include the inclusive production of 
jets\cite{DELPHI2,DELPHI3,GORDON1,KKLEINWORT} and single 
hadrons at large $p_T$\cite{GORDON2,BKK}. Applications 
at the $ep$ collider HERA include the production of 
jets and dijets \cite{KKLASEN,KRAMER,ZEUS1,ZEUS2,BUTTERWORTHKEK} and 
single hadrons\cite{GORDON2,BORZ,GRECO} at large $p_T$. Up to now, no 
contradiction with any data has been found. These pdfs have also been used 
in estimates for rarer processes such as the photoproduction of 
high-mass di-lepton pairs\cite{BCS} and large $p_T$ prompt 
photons\cite{GV,GSG}. 

At that time, and up until fairly recently, the only experimental 
information on the pdfs of the photon was obtained from studies of the 
structure function of the photon, $F^{\gamma}_{2}(x,Q^2)$, in
two photon collisions at $e^+e^-$ colliders. In LO QCD, 
$F^{\gamma}_{2}(x,Q^2)$ is related to the quark and anti-quark distributions 
in the photon, $q^{\gamma}_i(x,Q^2)$ and ${\overline {q}}^{\gamma}_i(x,Q^2)$
respectively, by,  
\begin{equation}
F^{\gamma}_{2}(x,Q^2)=\sum^{ N_f}_{i=1} xe_i^2\left( q^{\gamma}_i(x,Q^2)
+{\overline {q}}^{\gamma}_i(x,Q^2)\right)
\end{equation}
and thus the quark distributions are fairly directly determined from the
data on $F^{\gamma}_{2}(x,Q^2)$. Theoretically, this is a very interesting
area, because, at large $x$ and asymptotically large $Q^2$, the pdfs of
the photon and hence $F^{\gamma}_{2}(x,Q^2)$ are predicted from
PQCD in both LO\cite{Witten} and NLO\cite{BardeenBuras,FONPIL}. 
The so-called `anomalous' piece is {\bf calculable} as a function of 
$x$ and $Q^2$ and it is the $Q^2$ dependence (specifically $\propto \left( 
\alpha _s(Q^2)\right)^{-1}$) that ensures that it dominates 
at asymptotic values of $Q^2$. However, it is singular as $x\to 0$ and to 
regularise these singularities, some non-perturbative input to the 
Altarelli-Parisi(AP) equations is required, at a reference $Q^2=Q_0^2$. The 
anomalous $\left(\alpha_s(Q^2)\right)^{-1}$ behaviour of the pdfs 
arises because of the direct 
$\gamma \to q {\overline {q}}$ coupling, which gives inhomogeneous 
terms in the AP equations\cite{DEWITT}: this is the special feature of the 
photon. However, in the $Q^2$ range
experimentally accessible at present and in the foreseeable future, the effect 
of this input at a $Q^2=Q_0^2$ is significant, and so the analysis of the 
photon structure function is similar to that of a hadronic structure function.
Different groups make different choices of input scale $Q_0^2$, different 
forms of input pdfs, and all fit parameters to $F^{\gamma}_{2}(x,Q^2)$ data at 
various $Q^2$. The different possibilities have been recently reviewed in
ref\cite{SSV}.

The gluon distribution of the photon, $g^{\gamma}(x,Q^2)$, are not 
well constrained by these structure functions analyses\cite{GS}, 
essentially because the coupling of the AP equations 
for the gluon and singlet quark sectors is weak, and so the output
(evolved) quark pdfs and thus the evolved values of $F_2^\gamma (x,Q^2)$ 
(for $Q^2 \ge Q_0^2$) are rather independent of 
the input gluon distribution at $Q^2=Q_0^2$, $g^\gamma (x,Q_0^2)$. 
Furthermore, and even more importantly, unlike 
the case for hadrons,  $g^{\gamma}$ is not constrained by a momentum sum
rule\cite{GS,JKSLUND}. The result of all this is that the available 
parametrisations of the photon 
distributions agree reasonably well in their quark distributions but can 
have considerably different gluon distributions\cite{SSV,JKSLUND,VOGTLUND}. 
The input gluon 
distributions in the evolution equations, while not completely arbitrary, 
are currently just theoretically motivated guesses. 

Since our distributions were published, there has been much 
experimental activity in this and related areas. Not only has the 
amount of data on $F_2^\gamma $ been greatly increased, but 
data on jet production in both $\gamma \gamma $ collisions at $e^+e^-$ 
colliders and in $\gamma p $ collisions at HERA have appeared. These 
provide information on both quark and gluon distributions in the photon 
via {\bf resolved} photon processes\cite{DGOD}, as we shall see.

In view of this and also in view of the fact that there is expected 
to be an explosion of data in all of these areas in the next few years, it 
seems like an opportune moment to provide improved versions of our pdfs.
The new data will come from higher luminosity running at HERA and also the 
commissioning of LEP2, which is anticipated to provide several hundreds 
of inverse pb of integrated luminosity in $e^+e^-$ collisions over 
its lifetime.  The improvements we anticipate making are as follows: 

(1) Extending the kinematic range  of validity in $Q^2$ from 
$Q^2\geq 5.3$ GeV$^2$ down to $Q^2\geq 3$ GeV$^2$.

(2) Fitting to {\bf all} available data on $F^{\gamma}_2$ to obtain a 
better determination of the 

quark distributions. 

(3) Constraining the input gluon distribution by fitting the 
TRISTAN data on jet production in $\gamma \gamma $ collisions. We will
also see what can be learned from HERA 
data on jet production.  

\vskip 0.1in

The plan of this paper is as follows. In subsect. II.A we discuss general 
considerations about the pdfs of the photon. In subsect. II.B, we set up 
the parametrisation of the LO pdfs and in subsect. II.C we 
discuss our prescription for going over to the NLO 
parametrisation. In both cases we pay particular attention 
to the large $x$ behaviour. In sect. III we present our fits to the data, 
considering separately the data on $F_2^\gamma  $  (subsect. III.A), jets in 
$\gamma \gamma $ collisions (subsect. III.B), and jets in $\gamma p $ 
collisions (subsect. III.C). In sect. IV we discuss the properties of the 
new distributions and review future prospects for improving our 
knowledge of the photon pdfs. In sect. V we present conclusions.
 
\section{New Photon Distributions}

\subsection{General Considerations}

To obtain pdfs for the photon, we must start with input photon pdfs at some
reference scale $Q_0^2$, and choosing an appropriate value of $Q_0^2$ 
provides the first problem. Making a vector meson dominance (VMD) 
ansatz seems an obvious starting point as VMD provides a connection 
between the photon and $\rho $ meson pdfs. If we then 
use SU(6) to relate the $\rho $ pdfs to those of the pion, we can 
then use experimental constraints on the pion pdfs from Drell-Yan lepton
pair and prompt photon production data\cite{D-Y}. In such a VMD ansatz, 
the gluon is fixed by a momentum sum rule via a mesonic momentum sum
rule. However, a problem arises here: if we choose a plausible VMD 
scale, $Q_0^2 \simeq 1$ GeV$^2$, a pure VMD input is known to 
be insufficient to fit the data at higher $Q^2$\cite{GGR,JAN1,JAN2}. To
circumvent this problem two separate approaches have been adopted. The
first is to maintain the VMD approach but to start the AP evolution at 
a very low scale significantly below $Q_0 \simeq 1$
$GeV$\cite{GRV,AUR1,AUR2,SaS1}. The other is to take an input scale 
significantly above $Q_0 \simeq 1$ GeV and fit the quark pdfs to the
$F_2^\gamma $ data here, which essentially means supplementing the VMD 
input with a point-like component, seemingly naturally provided by the 
Born-Box diagram\cite{GS,SaS1,DG,LAC,WHIT}. Unfortunately there is no 
correspondingly natural choice for the gluon density and a guess must be
made. This is the approach that we will adopt here, as we did in
ref\cite{GS}, though we will be reducing from the value chosen there, 
$Q_0^2=5.3$ GeV$^2$, to $Q_0^2=3$ GeV$^2$. In ref\cite{GS} we 
advocated choosing the gluon content to ensure that the ratio 
of momentum carried by the quarks and antiquarks relative to 
that carried by the gluons, $R_2(Q^2)$, should lie between 1 and 3. The 
motivation for these limits is as follows. At low $Q^2$ we expect VMD to 
be good, the photon to be hadronic, and $R_2\simeq 1$ as for a hadron: at 
high $Q^2$, we 
expect the asymptotic result of PQCD\cite{FRAGUN}, that $R_2=99/32\simeq 3$ 
to hold, and so we expect a steady increase from 1 to 3 as $Q^2$ increases. 
It is interesting that at the time we noted that the LAC distributions 
have such a large gluon component that they are in danger of violating 
this bound. Since then LAC3 has been rejected by the data, because the 
large gluon component at large $x$ overestimates the jet cross section in both 
$\gamma \gamma $ and $\gamma p $ collisions and LAC1 seems to have too 
many gluons at small $x$ compared to the H1 data. We will expand on this later.

\subsection{Input Distributions: LO case}

To obtain regularised distributions in LO we separate the photon 
structure function into a hadronic and a point-like part\cite{GGR}
\begin{equation}
F^{\gamma}_2(x,Q^2_0)=F^{PL}_2(x,Q^2_0)+F^{HAD}_2(x,Q^2_0)
\end{equation}
The hadronic part, which we assume can be approximated by VMD, is important at
small values of $x$ ($\leq 0.2$) while the
point-like part, which we base on the Quark Parton Model (QPM) formula,
models the medium to large $x$ part.

The hadronic part of the input was chosen according to standard VMD ideas.
We assume that the hadronic photon can be represented by the $\rho^0$
meson
\begin{equation}
F^{\gamma}_{2,VMD}(x,Q^2)=\frac{4\pi
\alpha_{em}}{f^2_\rho} \sum_ie^2_ixq^{\rho^o}_i(x,Q^2) \label{eq:vmdd}
\end{equation}
where
\begin{math}
f^2_{\rho}/4{\pi}\approx 2.2\;
\end{math}
, and use SU(6) and isospin invariance to relate the $\rho^0$ distributions 
to those of the charged pions which are constrained by data\cite{D-Y}:
\begin{equation}
q^{\rho^o}_i(x,Q^2_0)=q^{\pi^0}_i(x)=\frac{1}{2}(q^{\pi^+}_i(x)
+q^{\pi^-}_i(x))
\end{equation}
The above assumptions are supported by the fact that the VMD estimates of the
structure function $F^{\gamma}_2(x,Q^2)$ base agree with the data
at $Q^2=1$ GeV$^2$\cite{TPC1,TPC2}. We also make the plausible assumption that 
the VMD breakdown into singlet, non-singlet and gluon sectors is also 
reliable. In terms of valence $v^{\pi}(x)$ and sea, $\zeta^\pi(x)$, 
these are given by
\begin{eqnarray}
q^{\gamma}_{NS,VMD}(x,Q^2_0)&=& \kappa \frac{4\pi \alpha_{em}}{f^2_\rho}
\left(\frac{1}{9}v^{\pi} \right) \nonumber \\
\Sigma^{\gamma}_{VMD}(x,Q^2_0)&=&\kappa \frac{4\pi
\alpha_{em}}{f^2_\rho}(2v^{\pi}+6\zeta^{\pi}) \\
   G^{\gamma}_{VMD}(x,Q^2_0)&=& \kappa \frac{4\pi
\alpha_{em}}{f^2_\rho}G^{\pi}(x) \nonumber
\end{eqnarray}
where we have used $N_f=3$ flavours since the heavy quark contributions are
expected to be very small at this $Q^2$.
The constant $\kappa$, where $1\leq \kappa \leq 2$, was introduced in
ref.\cite{GR} to represent the uncertainty over the inclusion of higher mass 
vector mesons. We take the following forms\cite{GS}:
\begin{eqnarray}
    xv^{\pi}(x)&=&A\sqrt{x}(1-x) \nonumber \\
x\zeta^{\pi}(x)&=&B(1-x)^5                 \\
    xG^{\pi}(x)&=&C(1-x)^3      \nonumber
\end{eqnarray}
The conservation of baryon number fixes $A$ at 0.75, fixing the momentum 
fraction of the hadronic part carried by the valence sector at $40\%$. The 
momentum fractions carried by the sea and gluon sectors are fixed by $B$ and 
$C$ which we use, along with $\kappa$ as free parameters in our fits 
to the $F_2^\gamma $ data.

For the point-like part we use the lowest order Bethe-Heitler (B-H) 
form\cite{BUDNEV,GOTS}
\begin{eqnarray}
q^{\gamma}_i(x,Q^2)&=&3e^2_i\frac{\alpha_{em}}{\pi}\left[\beta
\left( (8x(1-x)-1-\frac{4m^2_i}{Q^2}x(1-x)\right) \right. \nonumber \\
 & &\left. + \left( x^2+(1-x)^2+\frac{4m^2_i}{Q^2}x(1-3x)-
\frac{8m^4_i}{Q^4}x^2\right) \ln\left( \frac{1+\beta}{1-\beta}\right) \right]
\end{eqnarray}
for $\beta^2\geq 0$ where
\begin{displaymath}
\beta^2=1-\frac{4m^2_i x}{(1-x)Q^2}
\end{displaymath}
and $m_i$ are the light quark masses which are taken as fairly closely 
constrained parameters. Now $q^{\gamma}_i(x,Q^2)=0$ for $\beta^2 \le 0$, 
i.e. for 
\begin{equation} 
1\ge x \ge x_i = {{Q^2}\over{Q^2+4m_i^2}} 
\end{equation}
and typically $x_i\simeq 0.9$ for light quarks and $Q_0^2=3$ GeV$^2$. Also the 
VMD piece vanishes as $x\to 1$ as a power of $(1-x)$. Hence our quark 
distributions will be greatly suppressed at large $x$, a point we will 
return to later.
The parametric form for the distributions are thus
\begin{equation}
q_i^{\gamma}(x,Q^2_0)=q_i^{PL}(x,m_u,m_s)+q_i^{HAD}(x,\kappa,B,C)
\end{equation}
for the singlet and non-singlet sectors and for the gluon we take the form:
\begin{equation}
g^{\gamma}(x,Q_0^2)=g(x,\kappa,C)+\frac{2}{\beta_0}\lambda P_{gq}(x)*
\Sigma(x,m_i)
\end{equation}
and the convolution * is defined by 
\begin{equation}
A(x)*B(x)=\int^1_x\frac{dy}{y}A(x/y)B(y)
\end{equation}
and $\lambda $ is a parameter which enables us to adjust the component 
of the gluon input
estimated from bremsstrahlung off the singlet quarks. This will increase the 
fraction of the photon momentum carried by the gluons, and
compensate for the increase in the momentum carried by the
quarks due to inclusion of the point-like contribution. In our earlier 
work\cite{GS}, where we took $Q_0^2=5.3$ GeV$^2$, we presented two 
distributions, GS(I) and GS(II), corresponding to $\lambda = 1,0$ 
respectively: these numbers cannot be 
compared directly with those in the present work, as a different value of 
$Q_0^2$ is used. 

\subsection{ Input Distributions: NLO case}

The input distributions in NLO were obtained from the LO ones by 
demanding that the structure function, $F^{\gamma}_2(x,Q^2_0)$, be the same at
$Q_0^2=3$ GeV$^2$ in LO and NLO i.e.
\begin{equation}
F_2^{\gamma, NLO}(x,Q_0^2)=F_2^{\gamma ,LO}(x,Q_0^2)
\end{equation} 
in the same way as in ref.\cite{GS}. For the gluons we simply take 
\begin{equation}
g^{\gamma ,NLO}(x,Q_0^2)=g^{\gamma ,LO}(x,Q_0^2)
\end{equation}
but choosing the NLO quark distributions in order to satisfy the above 
condition is a non-trivial procedure which highlights the well-known fact that 
pdfs are renormalisation scheme dependent and have no physical significance 
on their own.

To proceed, we  assume that the distributions in NLO can be obtained 
from the LO ones by the addition of a correction term 
\begin{equation} 
q^{\gamma,
NLO}_i(x,Q^2_0)\ =\ q^{\gamma,LO}_i(x,Q^2_0)\ +\ \alpha_s(Q^2_0)
q^{(1)}_i(x,Q^2_0) \end{equation} 
These are substituted into the
defining equation for $F^{\gamma}_2(x,Q^2_0)$ which is 
\begin{eqnarray}
\lefteqn{\frac{1}{x} F^{\gamma}_2(x,Q^2)=\left( \delta(1-x)+
\frac{{\alpha}_s(Q^2)} {4\pi}B_q(x) \right)
{\ast}q^{\gamma}_{NS}(x,Q^2)} \nonumber \\ &
&+<e^2>\left(\delta(1-x)+\frac{{\alpha}_s(Q^2)}{4\pi}B_q(x) \right)
{\ast} {\Sigma}^{\gamma}(x,Q^2)   \\ &
&+<e^2>\frac{{\alpha}_s(Q^2)}{4\pi}B_G(x){\ast}
g^{\gamma}(x,Q^2)+{\delta}_{\gamma}B_{\gamma}(x),  \nonumber
\end{eqnarray} 
where 
\begin{equation}
\delta_{\gamma}=3N_f<e^4>\frac{\alpha_{em}}{4\pi}. \end{equation} 
and  the functions $B_q(x),\;B_G(x)$ and $B_{\gamma}(x)$ are the Wilson
coefficient functions. ${\alpha}_s(Q^2)$ is now the two loop coupling 
constant defined by 
\begin{equation} 
\alpha_s(Q^2)\approx
\frac{4\pi}{\beta_0\ln(Q^2/\Lambda^2)}\left[ 1-
\frac{\beta_1\ln[\ln(Q^2/\Lambda^2)]}{\beta_0^2\ln(Q^2/\Lambda^2)}\right].
\end{equation} 
Combining these equations at $Q^2=Q^2_0$ with a few extra assumptions to get 
the singlet and non-singlet distributions enables us to 
extract the $q_i^{(1)}$ 
and hence NLO quark distributions\cite{GS}. Of course, to carry this out 
we need to specify the renormalisation scheme as the Wilson 
coefficients and hence the pdfs are scheme dependent. It is only in 
combinations such as eq.(2.14) that they have physical significance. We 
work in the ${\overline {MS}}$ scheme: expressions for the Wilson 
coefficients in this scheme are given in ref\cite{JAN2}. 

There is a subtlety peculiar to the photon in the ${\overline {MS}}$ 
scheme which is discussed in detail in ref.\cite{VOGTLUND,AUR2}. It arises 
because $B_{\gamma }$ is divergent and negative as $x\to 1$ 
($B_\gamma \propto \ln(1-x)$). This means that if we put our LO input into 
eq.(2.14), we would get a negative $F_2^\gamma (x,Q_0^2)$ at large $x$, 
though the problem goes away at very large $Q^2$. The method 
outlined here eliminates this problem of negative values
of $F^{\gamma}_2(x,Q^2_0)$ at large $x$ due to the $B_{\gamma}(x)$ term in the
definition of the structure function. What we have done in imposing eq.(2.11)
is essentially to add a term to the quark distributions to compensate for the 
$B_{\gamma }$ term. The resulting distributions are steeply rising with
$x$ and are quite different from the LO ones, as we shall see.

\section{Fits to data}

\subsection{Structure Function Data}

Here we fit {\bf all} the data on  $F^{\gamma}_2$ with $Q^2 \ge 3
\;$GeV$^2$: we include data  from
PETRA\cite{PLUTO1,PLUTO2,TASSO,JADE1,JADE2,CELLO1,CELLO2},
 PEP\cite{TPC1,TPC2,TPC3}, TRISTAN\cite{AMY1,AMY2,TOPAZ1,VENUS}, and 
LEP1\cite{DELPHI1,OPAL}. The data cover a wide range of $Q^2$ and some 
have the charm subtracted out and some do not and so we must have a 
consistent policy over its treatment. For $Q_0^2 \le Q^2\le Q_1^2=50$
GeV$^2$, 
we use $N_f =3$ evolution and when we compare to data which includes charm 
we add to $F_2^\gamma$ the B-H estimate of charm as provided by eq.(2.6), 
taking $m_c=1.5$ GeV. For $Q^2 \ge Q_1^2$, all of the data have charm 
included. For these, we use a $N_f=4$ evolution which has been 
started from $Q^2 =Q_2^2=10$ GeV$^2$ with inputs generated by our $N_f=3$ 
evolution from $Q_0^2$ to $Q_2^2$ supplemented by the addition of the B-H 
term at $Q_2^2$. Throughout we keep $\Lambda _{{\overline {MS}}}$ 
fixed at 200 MeV.

In figs.(1-4) we show the fits to the data on the $x$-distributions 
at various $Q^2$ in both LO and NLO. To confirm the insensitivity of 
structure function analyses to the gluon content of the 
photon we show a sample of fits corresponding to different values of 
$\lambda $. It can be seen that the curves are indistinguishable, except 
at small $x$. 

In fig, 5 we show the fit to the data on 
\begin{equation}
\int _{0.3} ^{0.8} F_2^\gamma (x,Q^2)dx
\end{equation}
plotted against $\ln(Q^2)$ for LO and NLO. The two curves agree at
$Q_0^2$ but it can be argued that at higher $Q^2$, the NLO evolution
gives a somewhat better agreement with the data. It should also be noted
that our method of heavy flavour inclusion, discussed above, leads to no
discontinuity in the curves when one passes the heavy flavour
thresholds.

\subsection{Jets in $\gamma \gamma $ collisions}

Until relatively recently, studies of jets in $\gamma \gamma $ reactions 
were confined to model-dependent analyses of multiparticle 
distributions such as thrust. These were able to prove the existence\cite
{BACK,GGJETS} of the contribution from resolved photon processes\cite{DGODJ}, 
but were not precise enough to distinguish between different photon pdfs, 
though some effort was made along these lines by the later 
analyses\cite{DELPHI2,DELPHI3}, indicating some preference for the 
GS\cite{GS} pdfs. 

Since then data on the production of jets in %\gamma \gamma $ collisions has 
appeared from the TRISTAN collaborations AMY\cite{AMY3} and 
TOPAZ\cite{TOPAZ1,TOPAZ2}. Here we will attempt to constrain the input 
gluon distribution by fitting to their data on single and 
two-jet production. These have been shown to be somewhat sensitive to 
$g^{\gamma}$ in that they cannot be fitted with $g^\gamma =0$ and also have 
been used to rule out the somewhat extreme LAC3\cite{LAC} 
parametrisation, which, with its large gluon component at large $x$, gives 
much too high a cross section. Some calculations of jet 
production in $\gamma \gamma $ collisions in NLO have been carried out\cite
{GORDON1,KKLEINWORT,AURJAP}, which have mainly concentrated on the 
theoretical uncertainties, although some preliminary phenomenology has 
been done\cite{KKLEINWORT,AURJAP}. In view of the theoretical and 
experimental uncertainties we feel that at the moment it is only worth 
carrying out a LO comparison with the data. 

In this, for the photon flux we use the Equivalent Photon Approximation(EPA)
\cite{EPA}, with parameters fixed according to 
the experimental conditions. We keep our quark distributions fixed at 
the values we obtained in 
subsect. IIIA, and fit the data by varying the gluon content achieved by 
varying the parameter $\lambda $. We find good fits to the data with 
$\lambda $ in the range 0.90-1.0. We show our fits to the data in figs. 6 
and 7. To illustrate our sensitivity to the gluon we also show sample fits 
with $g^\gamma =0$ and those corresponding to $\lambda = 0\; {\rm and }\; 1$

\subsection{Jets in $\gamma p$ Collisions}

At HERA, $ep$ interactions in the untagged case are dominated by 
photoproduction with the spectrum of the (mainly) real photons given by the 
EPA\cite{EPA}. The first data on jets\cite{HERAOLD} were superseded by the 
first jet cross section data which enabled the direct photon reactions 
to be separated from the resolved photon reactions\cite{ZEUS0} and 
for the LAC3\cite{LAC} pdfs to ruled out as they overestimated 
the jet cross section\cite{H11} because of their large gluon distribution. 
Since then much better quality data on jet and di-jet cross 
sections\cite{ZEUS1,ZEUS2,H12,H13} have been published which it is 
hoped can be used to give information on the pdfs of the photon. These data 
have been confronted by NLO calculations\cite{AFG2,KKS,KKLASEN} and the fits 
are not good\cite{KRAMER}. What conclusions can be drawn from this is not 
clear at the moment as there are other aspects of the data which are 
difficult to understand, for example, the energy flow out of the jet 
for the resolved component\cite{MAXFIELD}. Multiple parton 
interactions have been suggested as 
the reason\cite{MULT} but it is not clear how they can be incorporated into 
a NLO calculation. In a LO calculation, they give an effect in the right 
direction, increasing the cross section in the regions which are dominated by 
the resolved contribution, particularly at lower $p_T$, and they also 
help with the energy flow problem.

In view of this confusion, it seems premature to compare the results of a full 
NLO calculation with the data: we will simply confine ourselves to a few 
semi-quantitative remarks on what has been learned from the data that is 
relevant to pdfs. The demise of LAC3 shows that the we cannot have a large 
gluon distribution at large $x$. There are two other places where something 
relevant can be learnt. 

The first is the fact that many 
pdfs overestimate the photoproduction cross section in the region of 
extreme negative rapidity. This is true for both single 
jet\cite{AFG2,KKS,KRAMER} and dijet\cite{KKLASEN,KRAMER} cross sections. Here 
the cross section is supposed to be 
dominated by the direct contribution which depend only on the proton pdfs 
which are well-known from deep inelastic scattering from protons. In direct 
photon processes multiple parton interactions cannot be 
the culprit: in fact the energy flow is well understood here. It turns out 
that with the NLO pdfs used initially\cite{KKS,KRAMER,AFG2,KKS}, GRV and AFG, 
the resolved contributions make a 
significant contribution here, presumably because their quark pdfs are large 
at large $x$, and a parton with large $x$ contributes like a direct photon, 
as it brings all of the photon energy into the hard subprocess. This problem 
goes away if the GS distributions are used both in the dijet 
case\cite{KKLASEN,KRAMER} and in the single jet case\cite{KRAMER}. This is 
 presumably because they are 
relatively suppressed at large $x$, as we pointed out earlier, certainly in 
LO where the pdfs have physical significance, and this relative suppession is 
reflected in the NLO quark distributions\cite{SSV}. This difference in the 
quark distributions does not show up in structure function analyses: there 
are no data at large $x$ and are unlikely to be in the future as large $x$ 
corresponds to $W$, the energy of the $\gamma \gamma \to {\rm hadrons}$, 
process being very small. 

The other place is the gluon distribution of the photon which 
has been ``extracted'' by H1\cite{H13}, They used the following procedure: they 
constructed an $x_\gamma $ distribution of the dijet cross section, 
subtracted the contribution from direct photons to obtain 
the resolved cross section, and then subtracted the quark contribution from 
that, leaving the gluon distribution of the photon, which we plot 
against ours in fig. 8. There 
are many uncertainties associated with this procedure: it is only defined in LO
, scale dependence is obviously a problem, as are multiple 
interactions, which are larger at small $x_\gamma$. Also the quark 
contribution subtracted was that corresponding to GRV: this is 
probably irrelevant in view of the other uncertainties catalogued above. 
However, the result is perhaps indicative that the data 
prefer more moderate gluon distributions such as ours and GRV. It is 
the extreme gluons such as LAC1 and LAC3 that have problems.

\section{Discussion}

The new GS photon distributions are many respects similar to the GS1 set
of ref\cite{GS}. As we have discussed, the main improvements are; a
lower starting scale for the evolution and more importantly, we have
fitted to all available data on the photon structure function as well as
attempted to constrain the gluon input using single and two jet data
from TRISTAN. The HERA data on single and two jet production has also
provided some indication that smaller quark distributions such as ours are 
preferred at large-$x$ and not the large distributions given by most of
the other parametrisations on the market. We have attempted here to use
information from all possible sources, both recent and older,
experimental and theoretical, in order to constrain the parton distributions of 
the photon. For example the ratio $R_2(Q^2)$, discussed in the introduction,
which is the ratio of the momenta carried by quarks to gluons starts at
around $1$ and approaches $3$ as $Q^2$ increases. We regard this as a
kind of theoretical constraint on the gluon distributions. 

There are some areas where some improvements could be made in our
parametrisations. Two such areas are the flavour decomposition of the
structure function, and continuity in $Q^2$ of the quark distributions.
In our treatment, the four flavour evolution is started at $Q^2=10$
GeV$^2$ but not used except above $Q^2=50$ GeV$^2$. At this point the charm
distribution is taken as equal to that of the $u$-quark, which
undoubtedly overestimates it by a considerable amount, but
correspondingly, the $u$-quark distribution is reduced at this point,
rendering it discontinuous there. The sum of the quark
distributions, and hence $F_2^\gamma(x,Q^2)$ on the other hand has no 
discontinuity at this point. We therefore recommend that single flavour
distributions, such as required for charm production in a massless
framework not be calculated using our distributions. 

As mentioned above, although there are no data on $F^\gamma_2(x,Q^2)$   
at large-$x$, above $x\sim 0.8$, the recent HERA data on jet production
appears to prefer smaller quark distributions in this region. At
small-$x$, below $x\sim 0.05$, on the other hand we have no experimental 
information and hence cannot constrain the distributions here, except
via the theoretically motivated models. It is hoped that at in the near
future this situation may be remedied.

Our knowledge of the photon distributions has increased very
significantly over the past few years, thanks to new experimental data
from TRISTAN, LEP and HERA. The future outlook is even more optimistic,
as HERA is expected to run at higher luminousity and LEP2 will soon
start taking data. We anticipate that processes such as prompt photon
\cite{JKGOR}, heavy flavour and lepton pair production will be measured with 
good statistics at these machines, allowing tighter constraints on the
quark and gluon distributions of the photon.

\section{Conclusions}

We have attempted to update the GS photon distribution functions by, among 
other things, constraining the input gluon distributions by fitting to 
TRISTAN jet data. We find that although the data is useful as an indicator 
of the existence of $g^{\gamma}$ and to rule out extreme distributions, it 
is not good enough to fix the distribution better than current VMD 
estimates, or other reasonable fits. 

\section{ACKNOWLEDGEMENTS}
This work was supported in part by the U.S. Department of Energy,
Division of High Energy Physics, Contract W-31-109-ENG-38.

\newpage
\noindent
\begin{flushleft}
{\Large \bf Figure Captions}
\end{flushleft}

\begin{description}
\item[{\bf Fig. 1:}] Fits to PETRA data on the photon stucture function.
\item[{\bf Fig. 2:}] Fits to PEP data on the photon structure function. 
\item[{\bf Fig. 3:}] Fits to TRISTAN  data on the photon structure
function. At $Q^2=390$ GeV$^2$, in the second graph the solid curve is for the 
choice $\lambda=2.0$ (see text).
\item[{\bf Fig. 4:}] Fits to LEP1 data on the photon structure function.
At $Q^2=12$ GeV$^2$, in the second graph, the dot-dashed curve is for
the choice $\lambda=0$.
\item[{\bf Fig. 5:}] $Q^2$ dependence of the structure function
integrated over the region $0.3\leq x \leq 0.8$.
\item[{\bf Fig. 6:}] Fits to AMY data on $\gamma \gamma \to {\rm jets}$
for two choices of $\lambda$ and for $g^\gamma=0$; (a) fits to the $p_T$ 
distribution of the single-jet data (b) fits to the rapidity
distribution of the single-jet data and (c) fits to the $p_T$
distribution of the two-jet data. 
\item[{\bf Fig. 7;}] Fits to TOPAZ (a) single-jet and (b) two-jet data.
\item[{\bf Fig. 8;}] Comparison of the GS(HO), GS(LO) and GS(I) and
GS(II) LO \cite{GS} gluon distributions with the gluon distribution
extracted by the H1 Collaboration \cite{H13}.    
\end{description}

\begin{figure}
{\hskip 0.2cm}\hbox{\epsfxsize7.5cm\epsffile{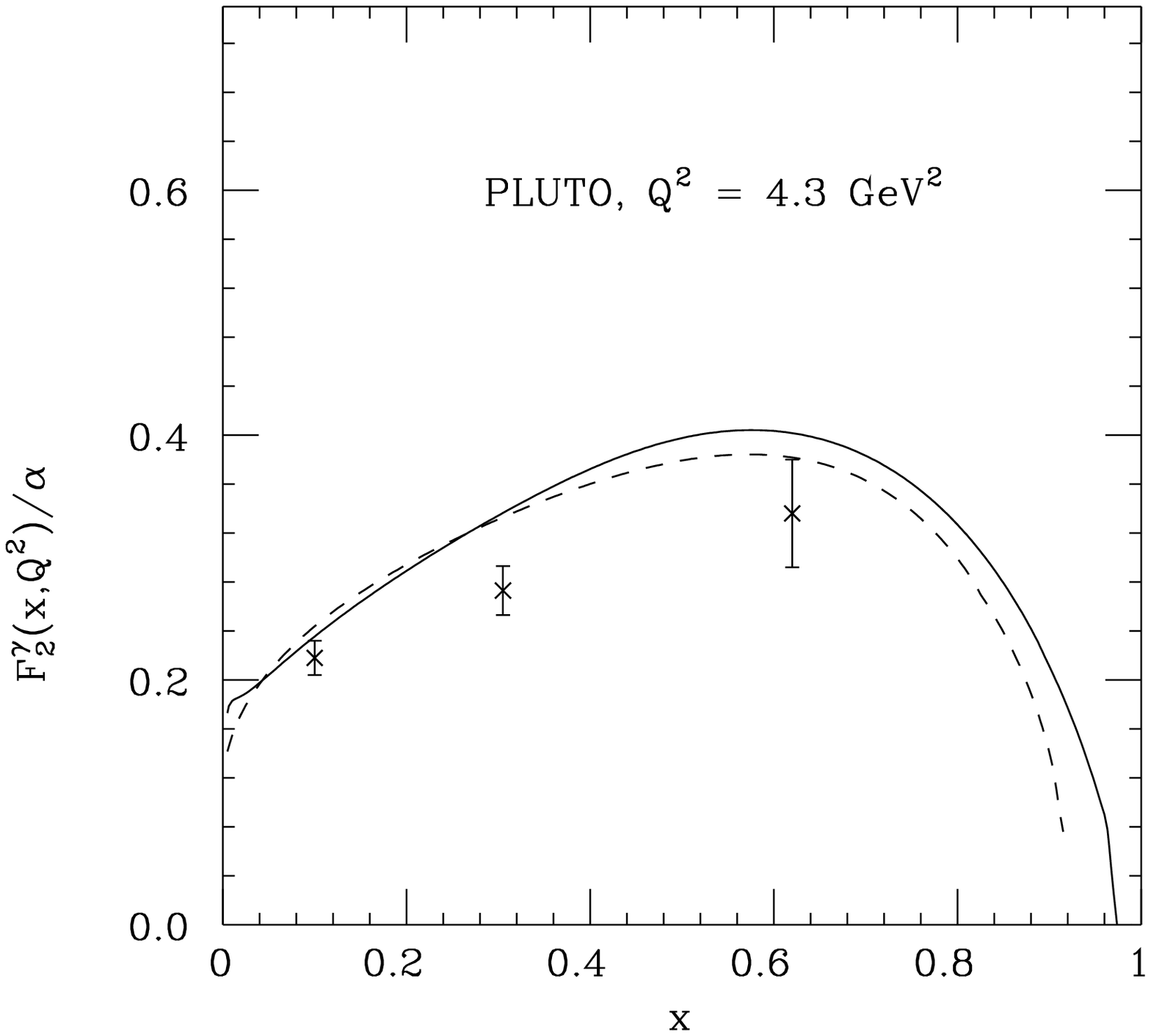}{\hskip 0.2cm}
\epsfxsize7.5cm\epsffile{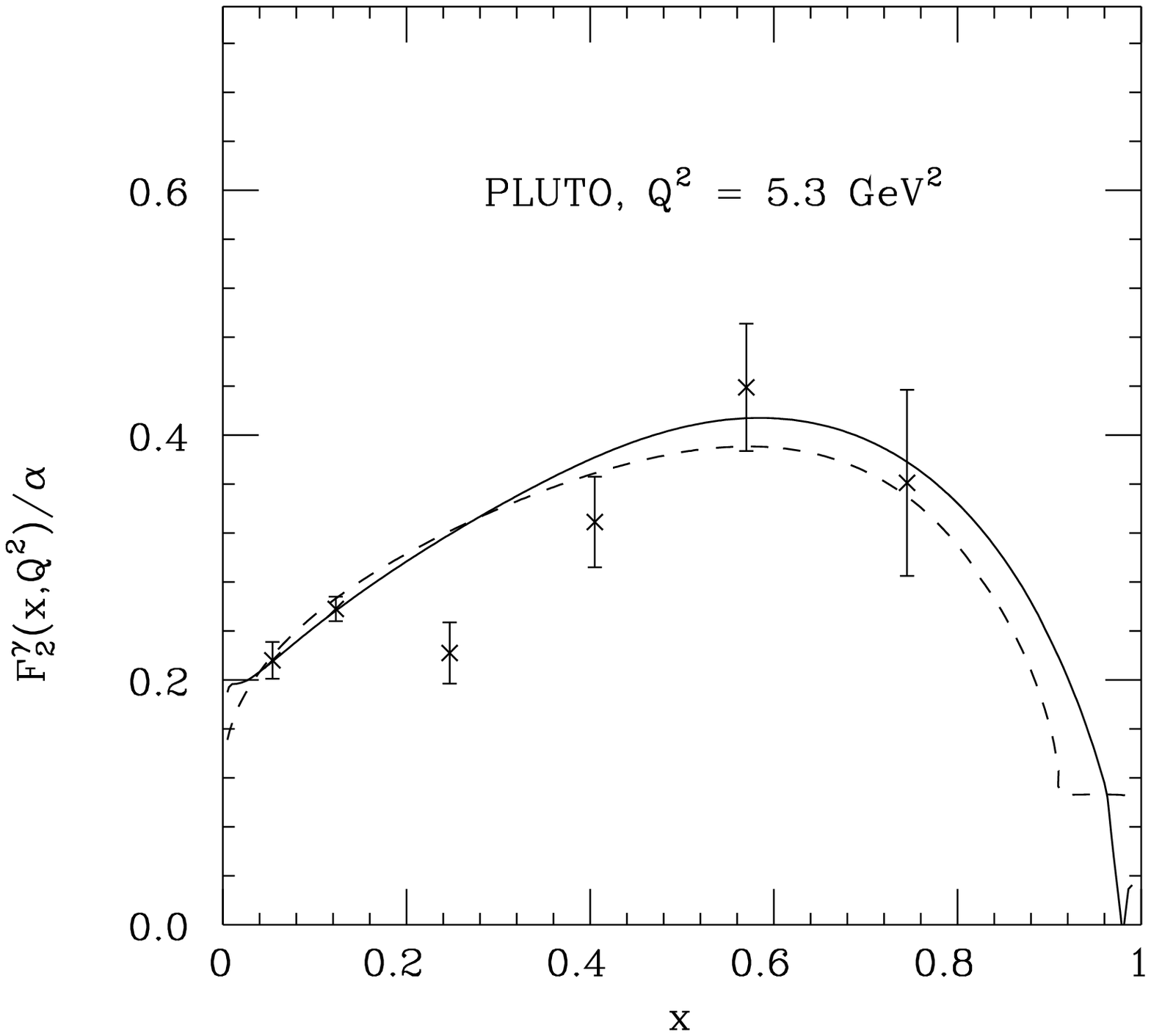}}
\end{figure}
\begin{figure}
{\hskip 0.2cm}\hbox{\epsfxsize7.5cm\epsffile{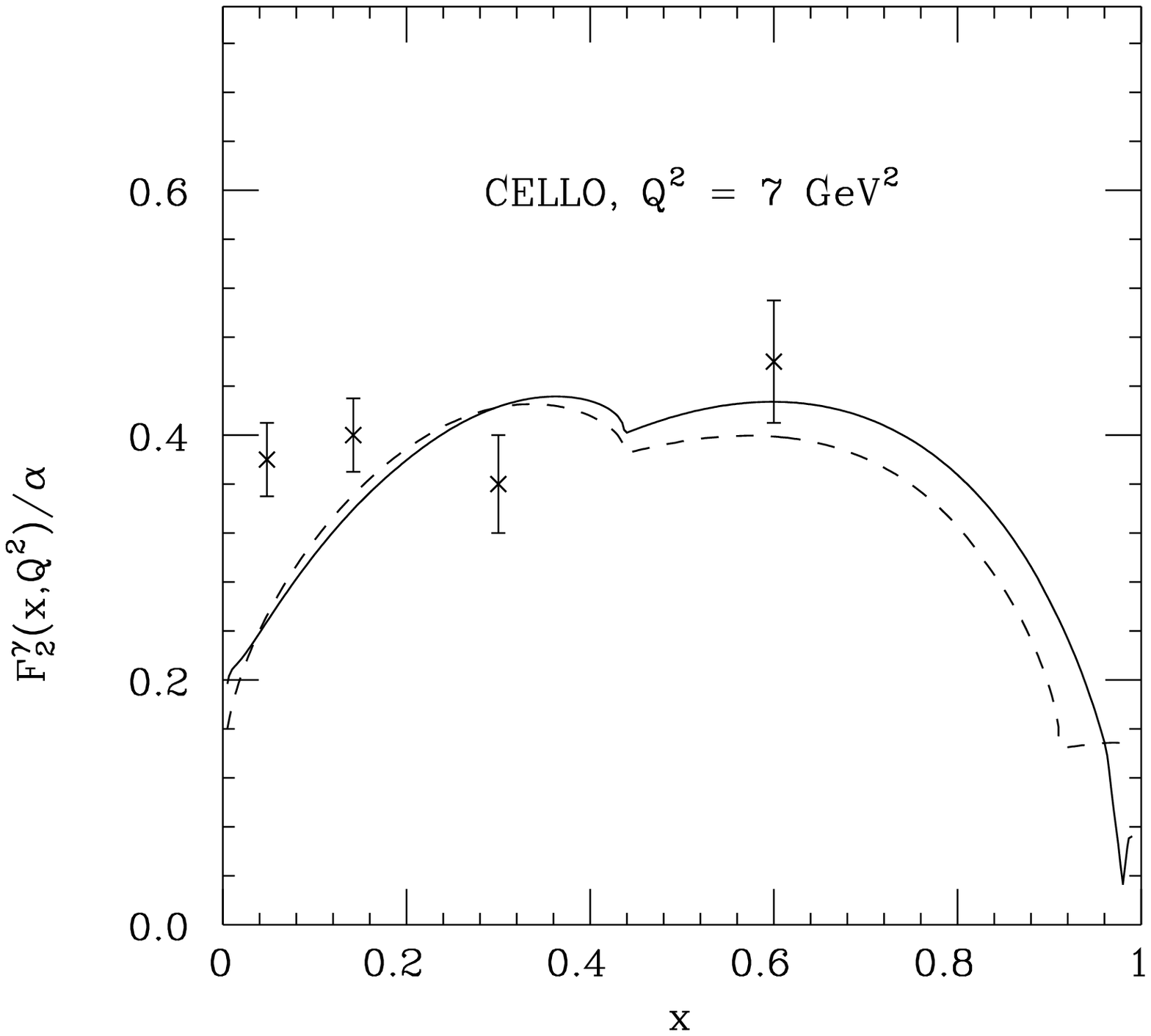}{\hskip 0.2cm}
\epsfxsize7.5cm\epsffile{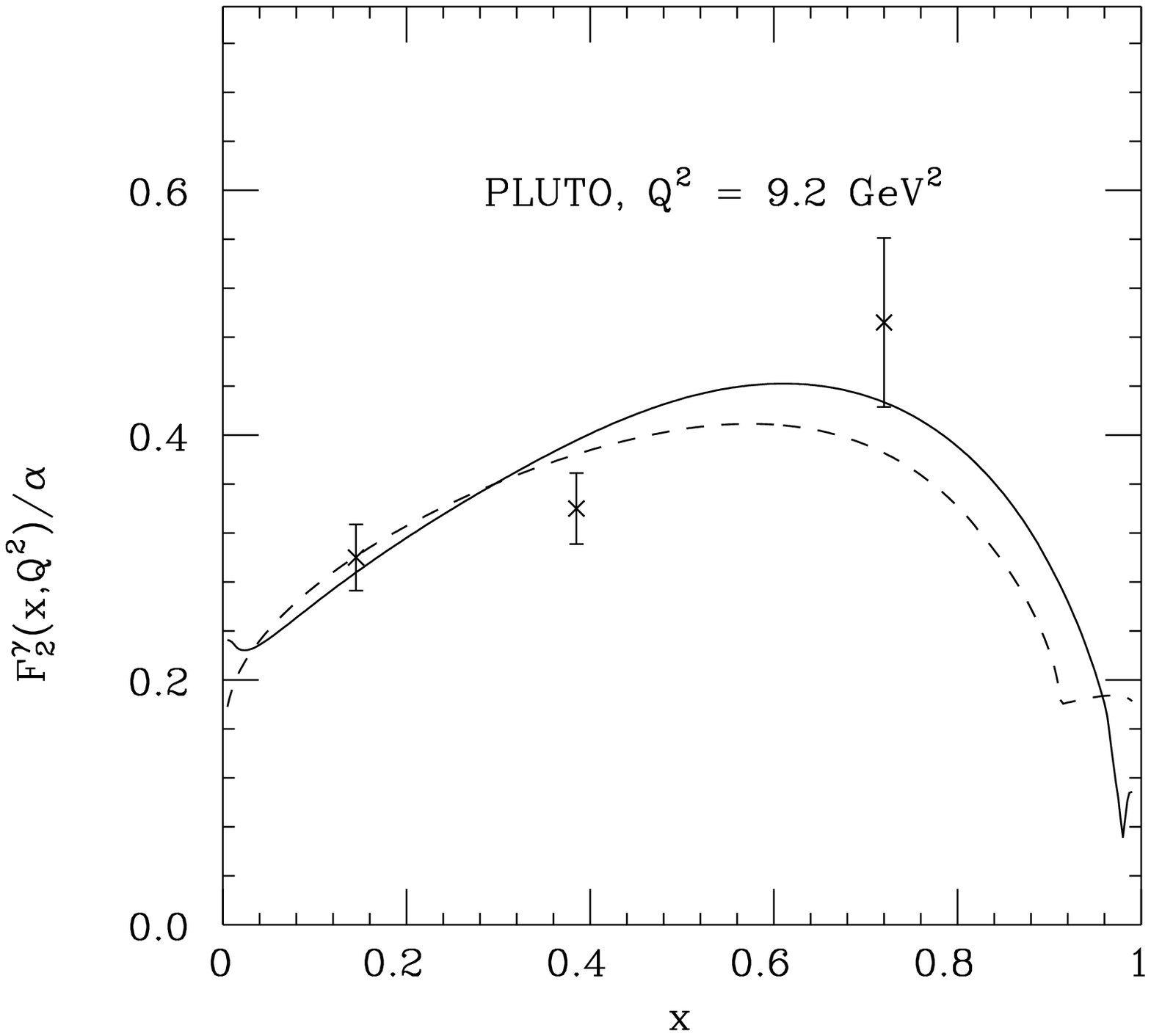}}
\end{figure}
\begin{figure}
{\hskip 0.2cm}\hbox{\epsfxsize7.5cm\epsffile{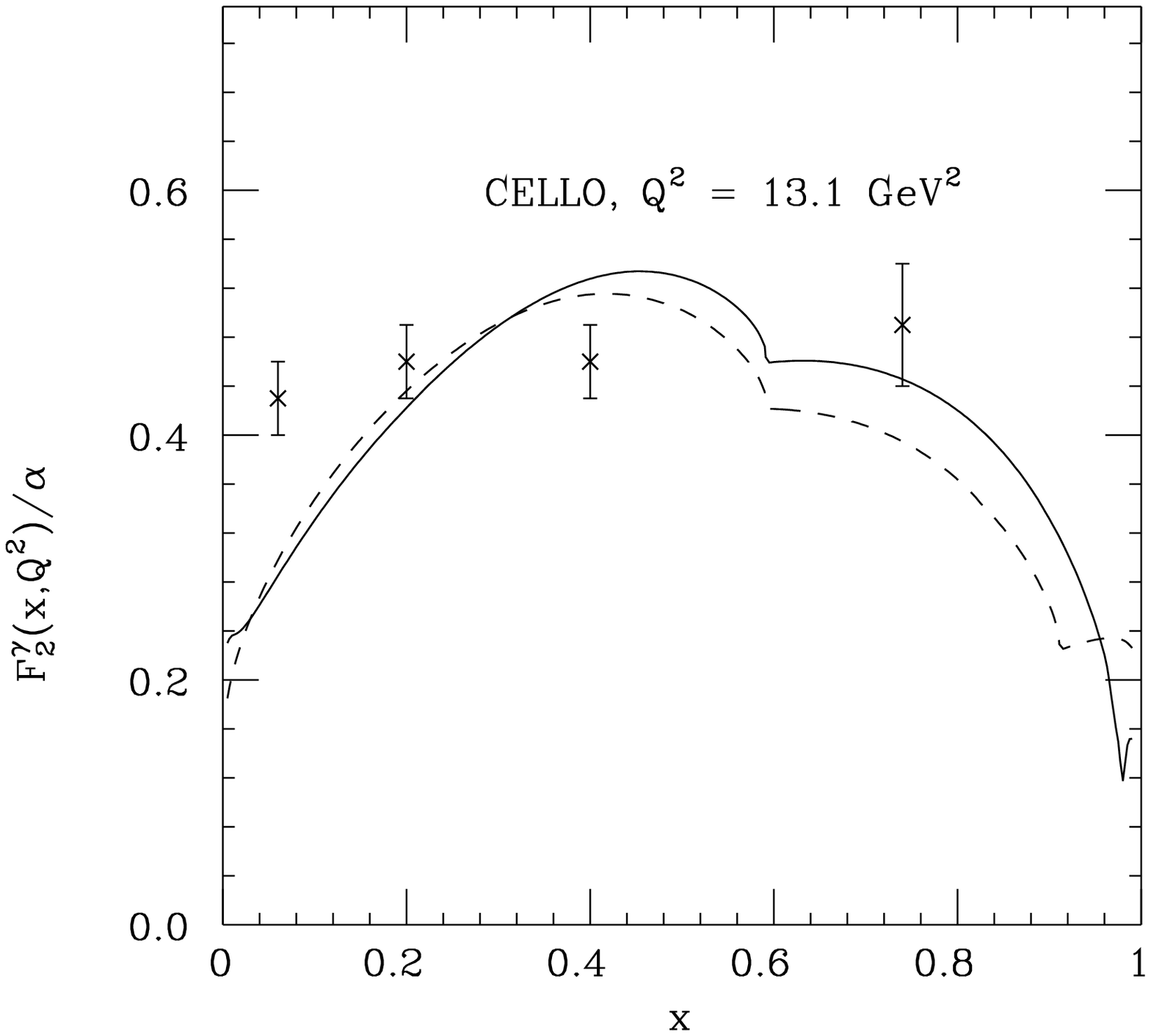}{\hskip 0.2cm}
\epsfxsize7.5cm\epsffile{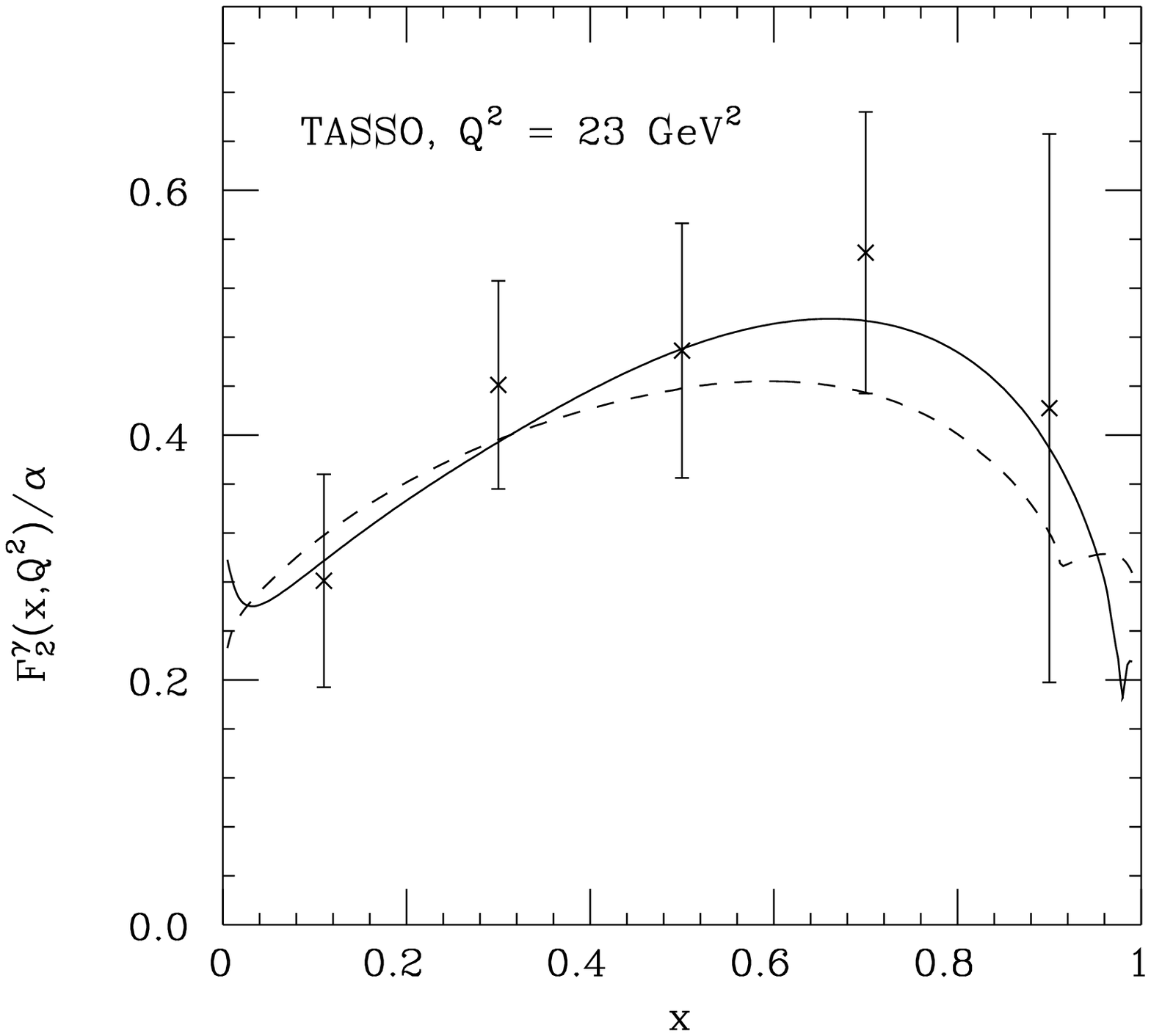}}
\end{figure}
\begin{figure}
{\hskip 0.2cm}\hbox{\epsfxsize7.5cm\epsffile{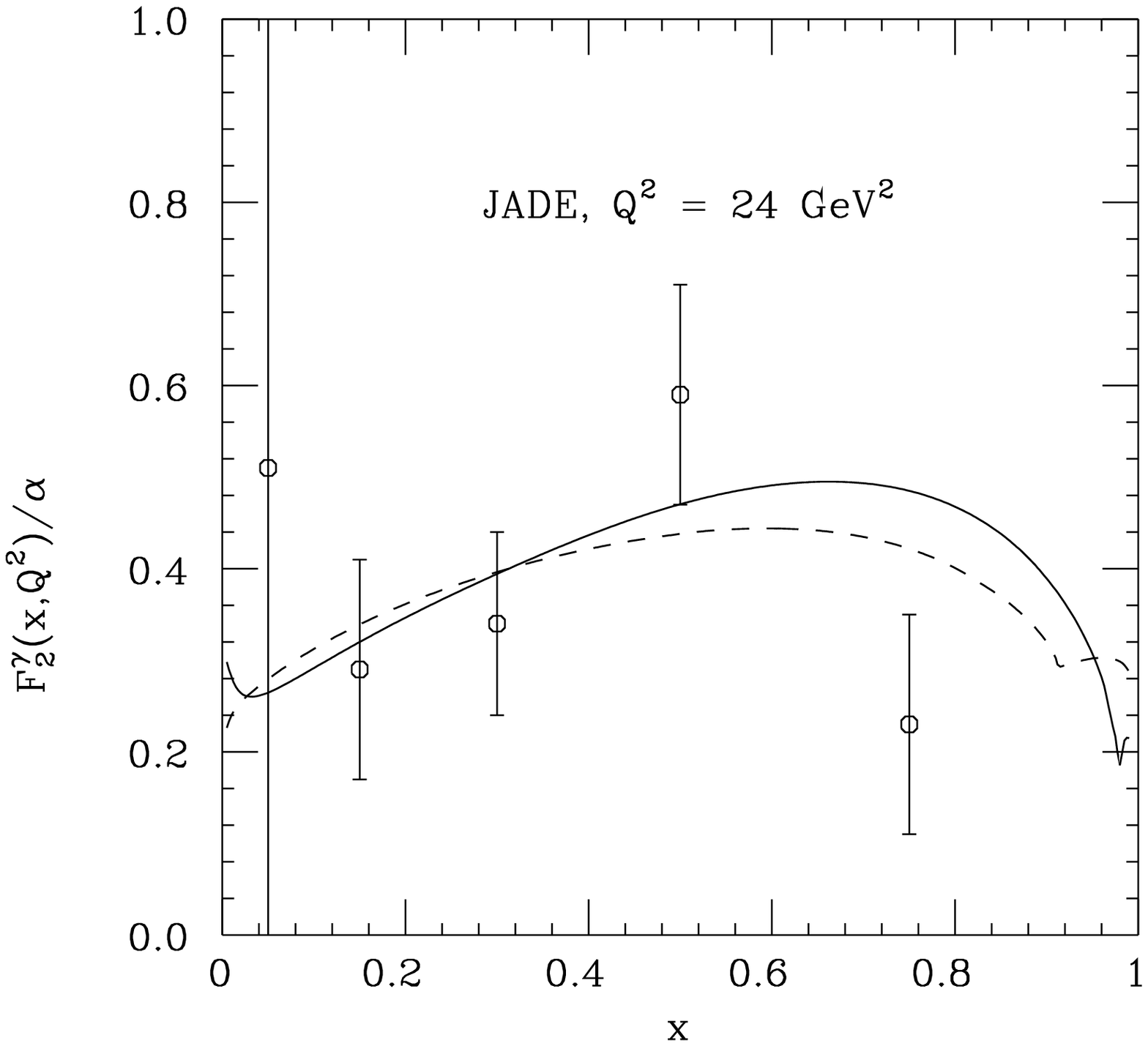}{\hskip 0.2cm}
\epsfxsize7.5cm\epsffile{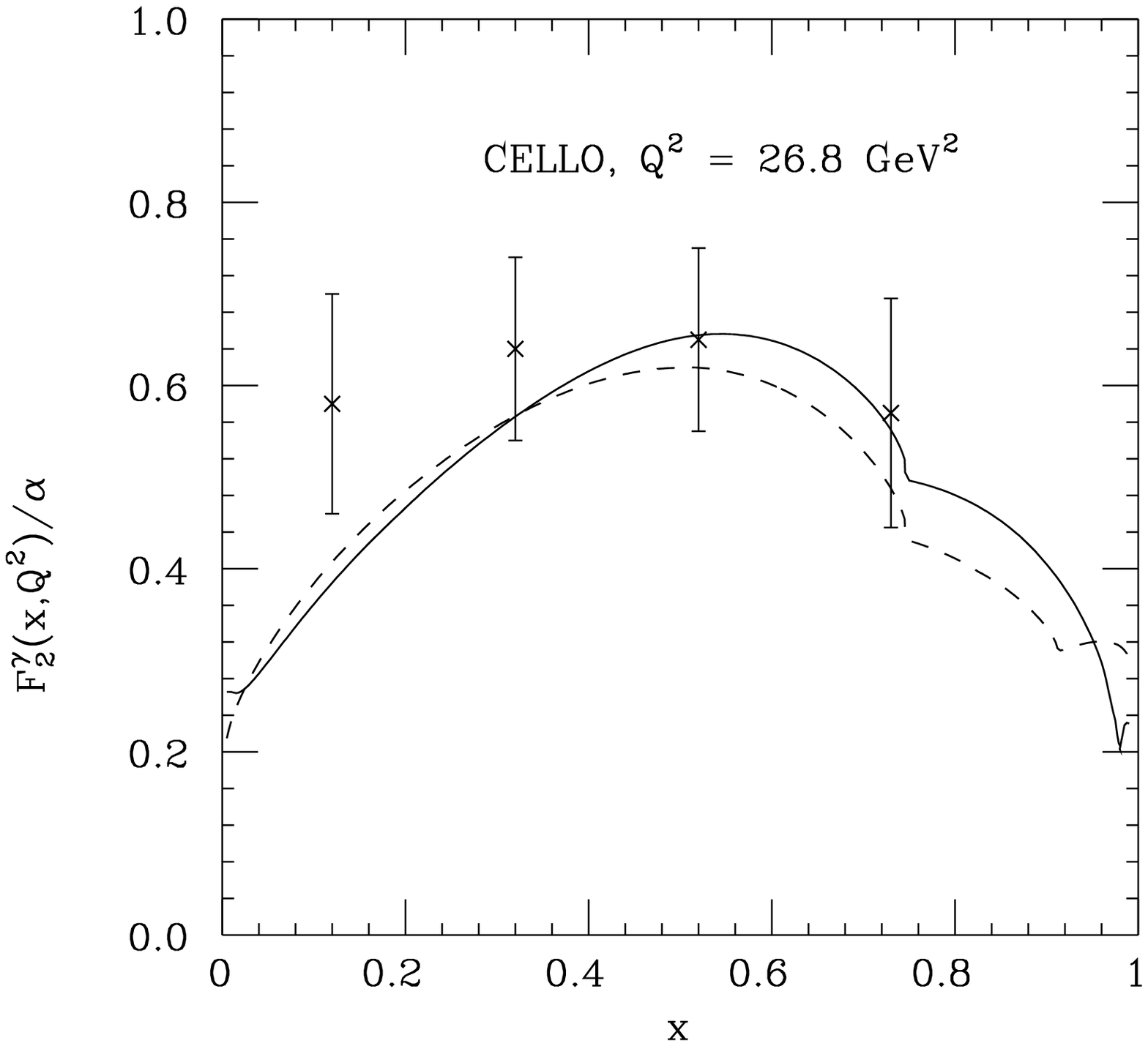}}
\end{figure}
\begin{figure}
{\hskip 0.2cm}\hbox{\epsfxsize7.5cm\epsffile{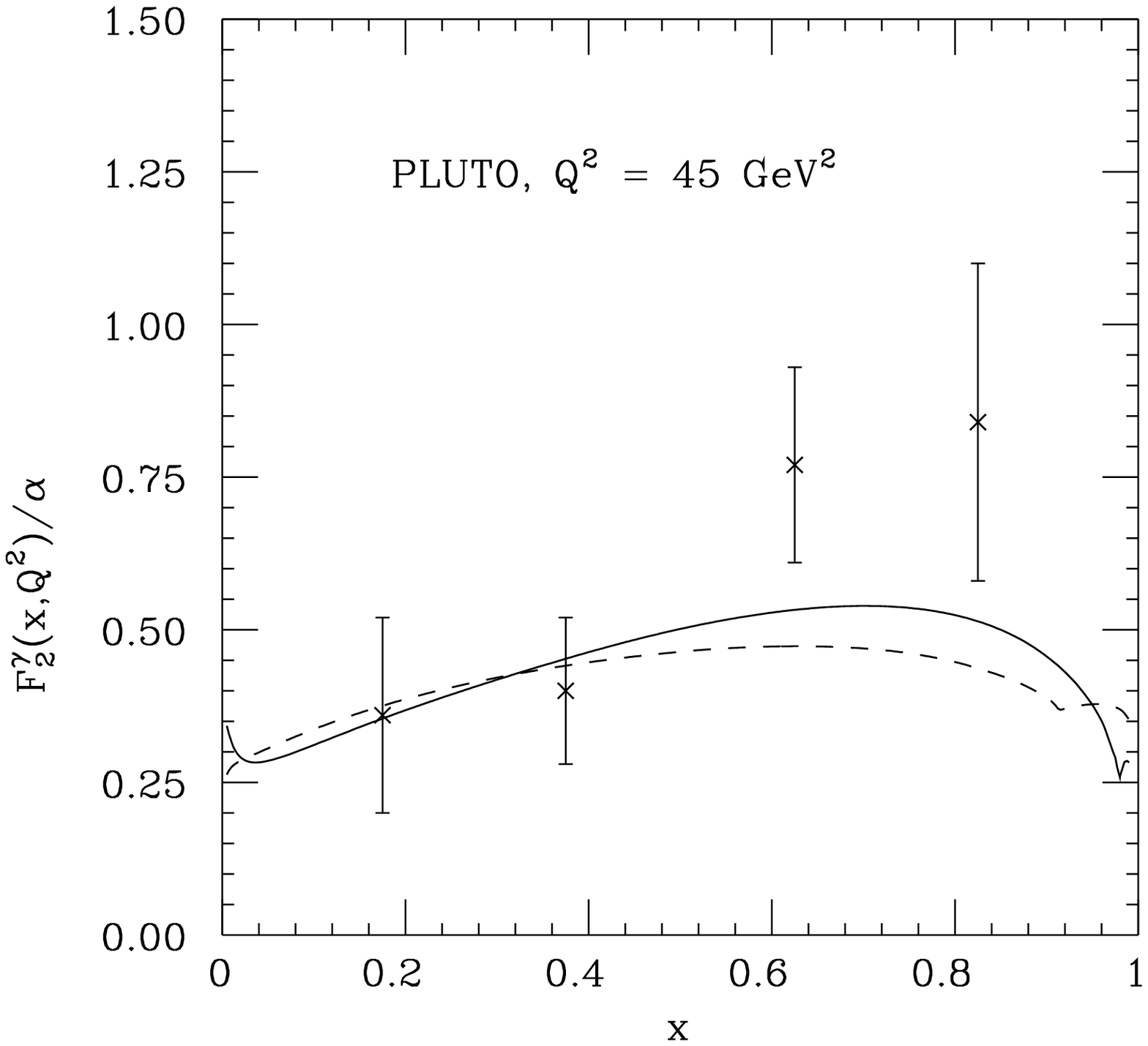}{\hskip 0.2cm}
\epsfxsize7.5cm\epsffile{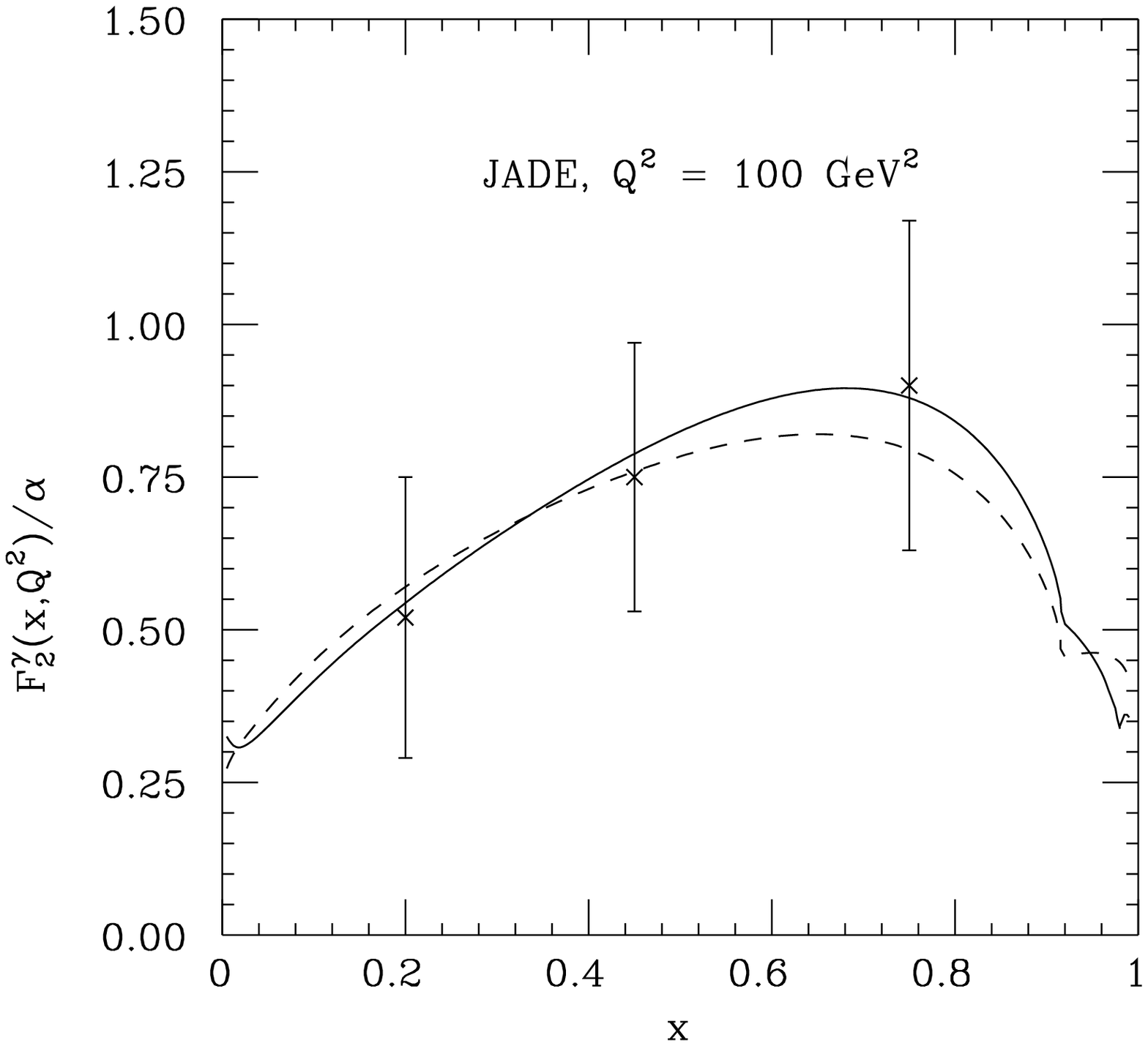}}
\fcaption{}
\end{figure}
%----------------------------------------------------------------
\pagebreak
\begin{figure}
{\hskip 0.2cm}\hbox{\epsfxsize7.5cm\epsffile{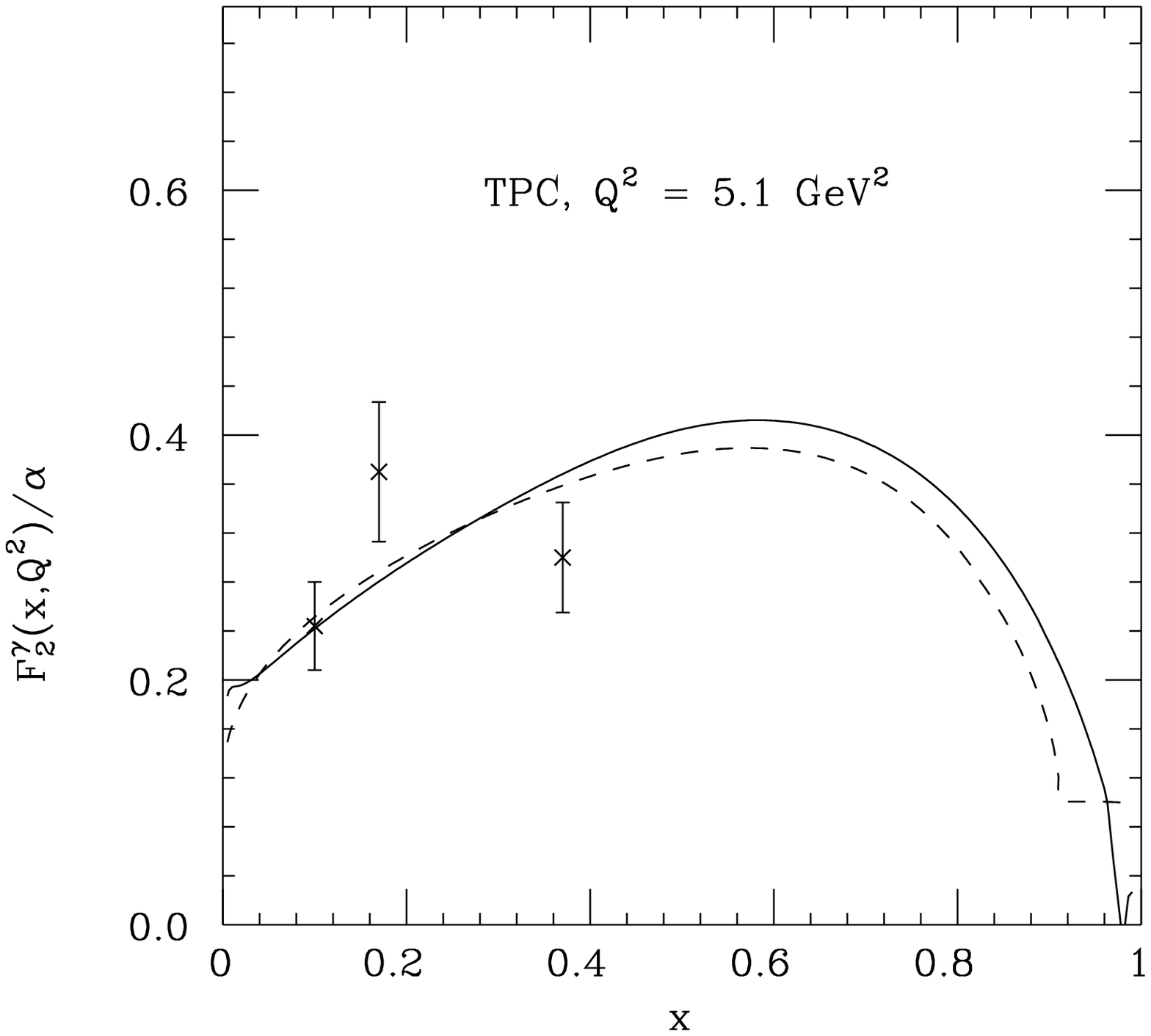}{\hskip 0.2cm}
\epsfxsize7.5cm\epsffile{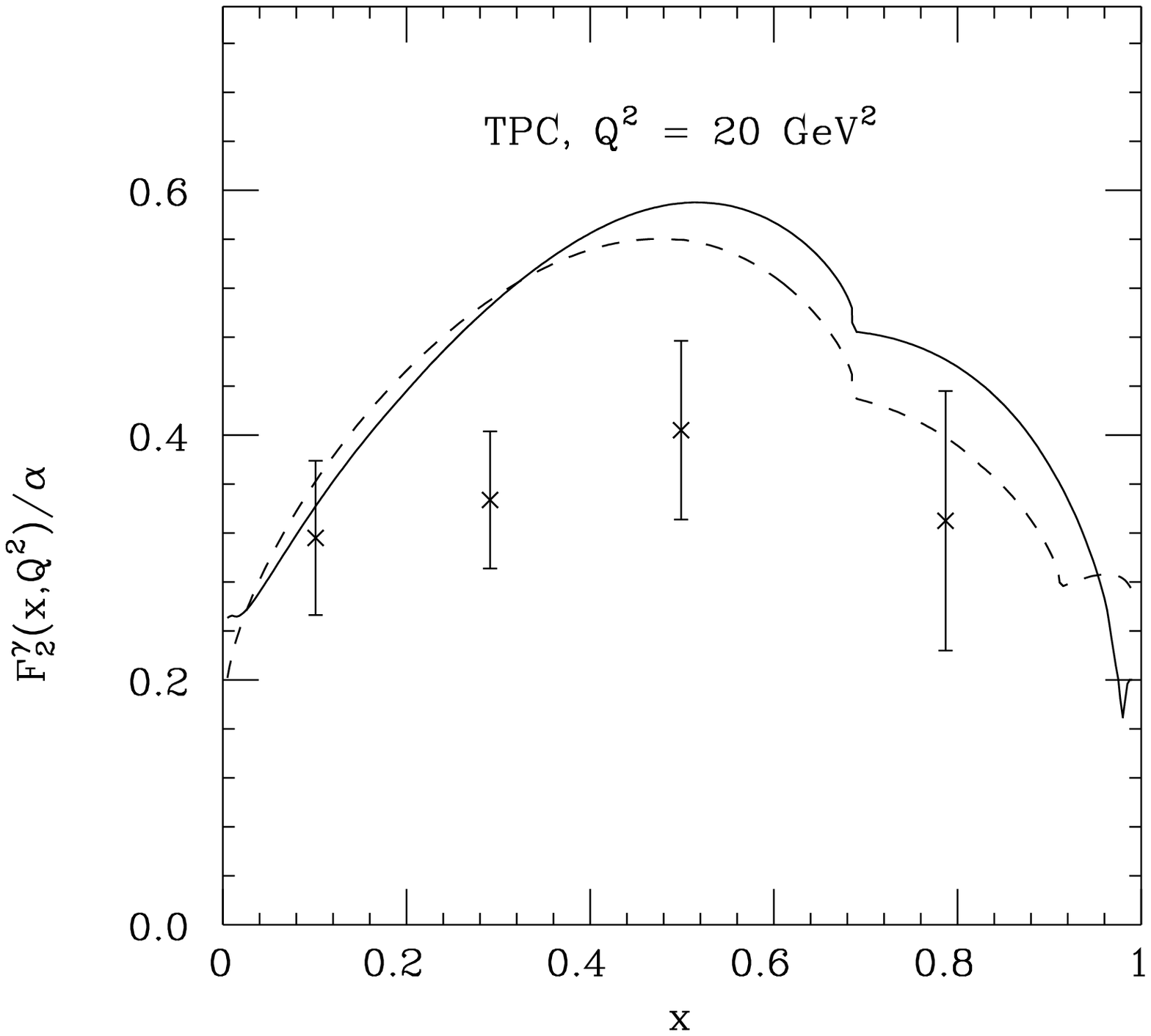}}
\fcaption{}
\end{figure}
%----------------------------------------------------------------
\pagebreak
\begin{figure}
{\hskip 0.2cm}\hbox{\epsfxsize7.5cm\epsffile{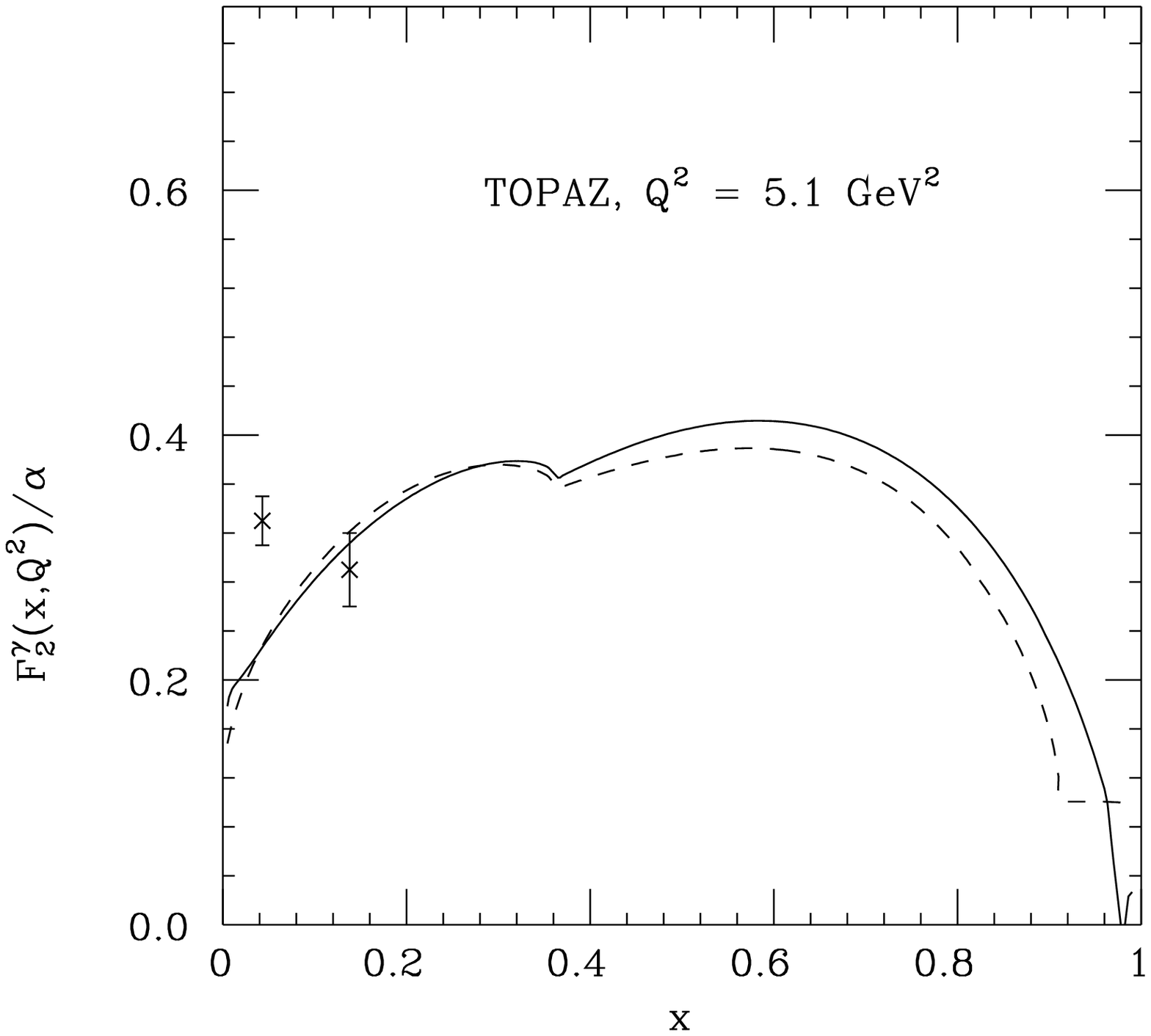}{\hskip 0.2cm}
\epsfxsize7.5cm\epsffile{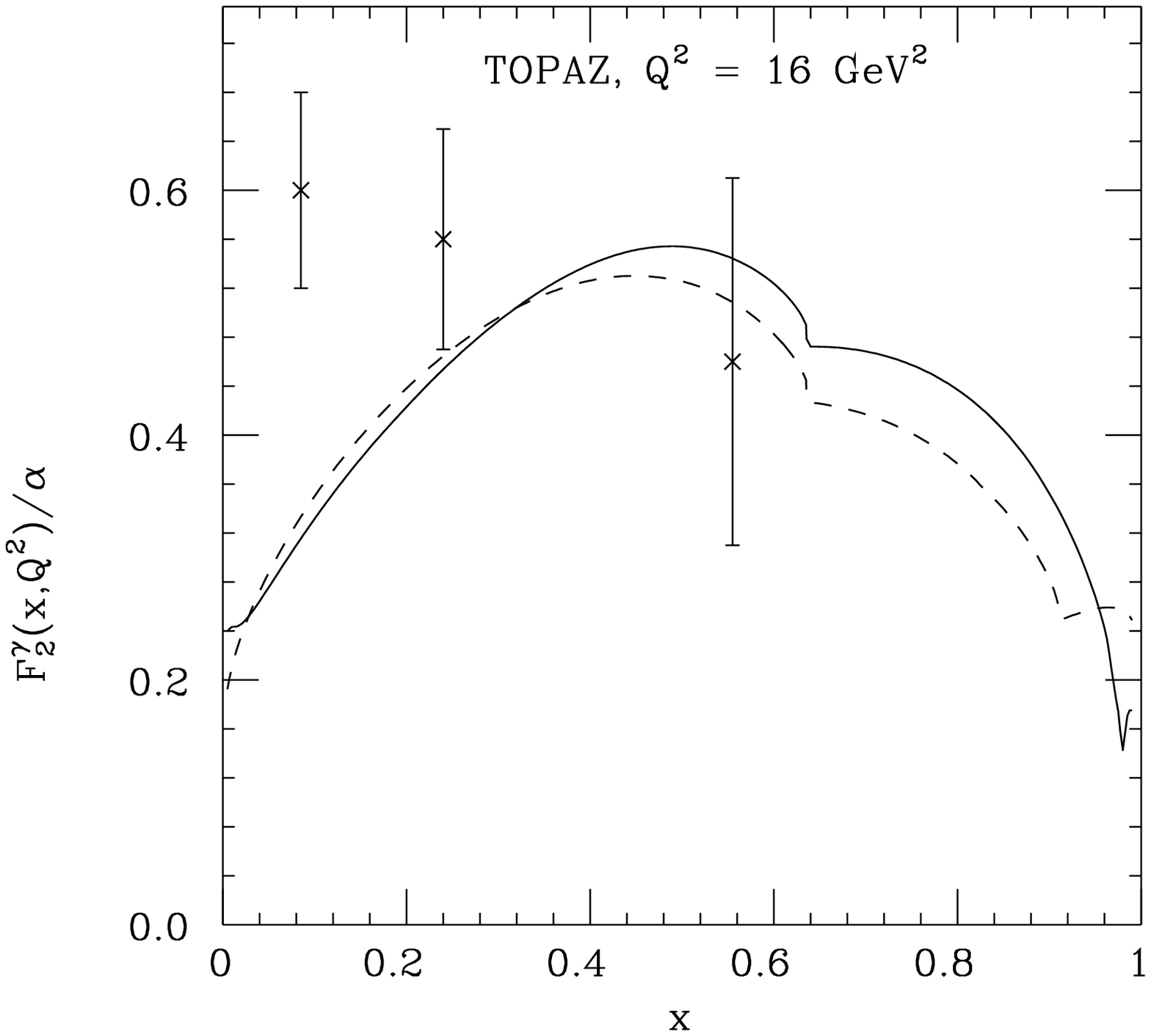}}
\end{figure}
\begin{figure}
{\hskip 0.2cm}\hbox{\epsfxsize7.5cm\epsffile{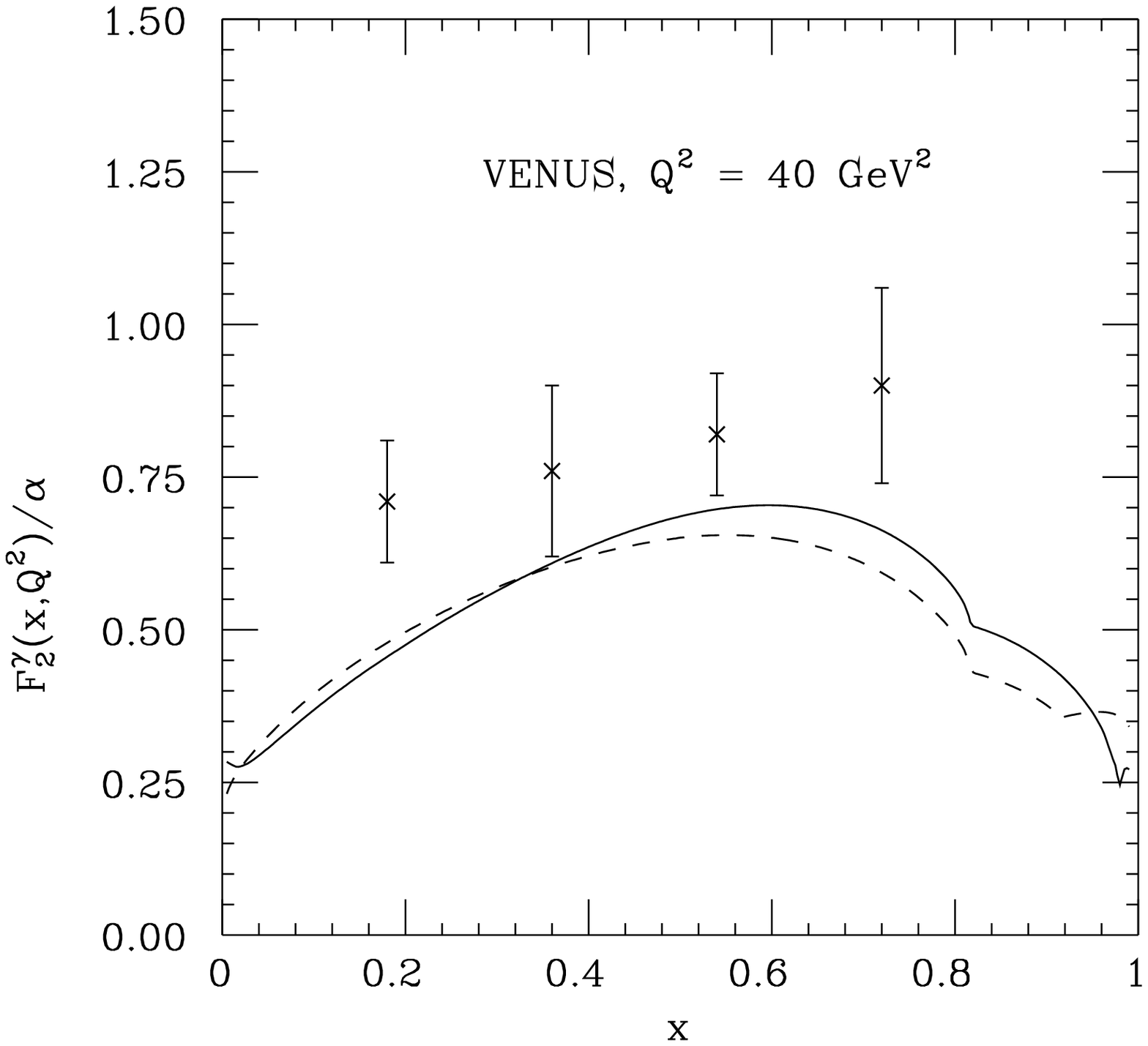}{\hskip 0.2cm}
\epsfxsize7.5cm\epsffile{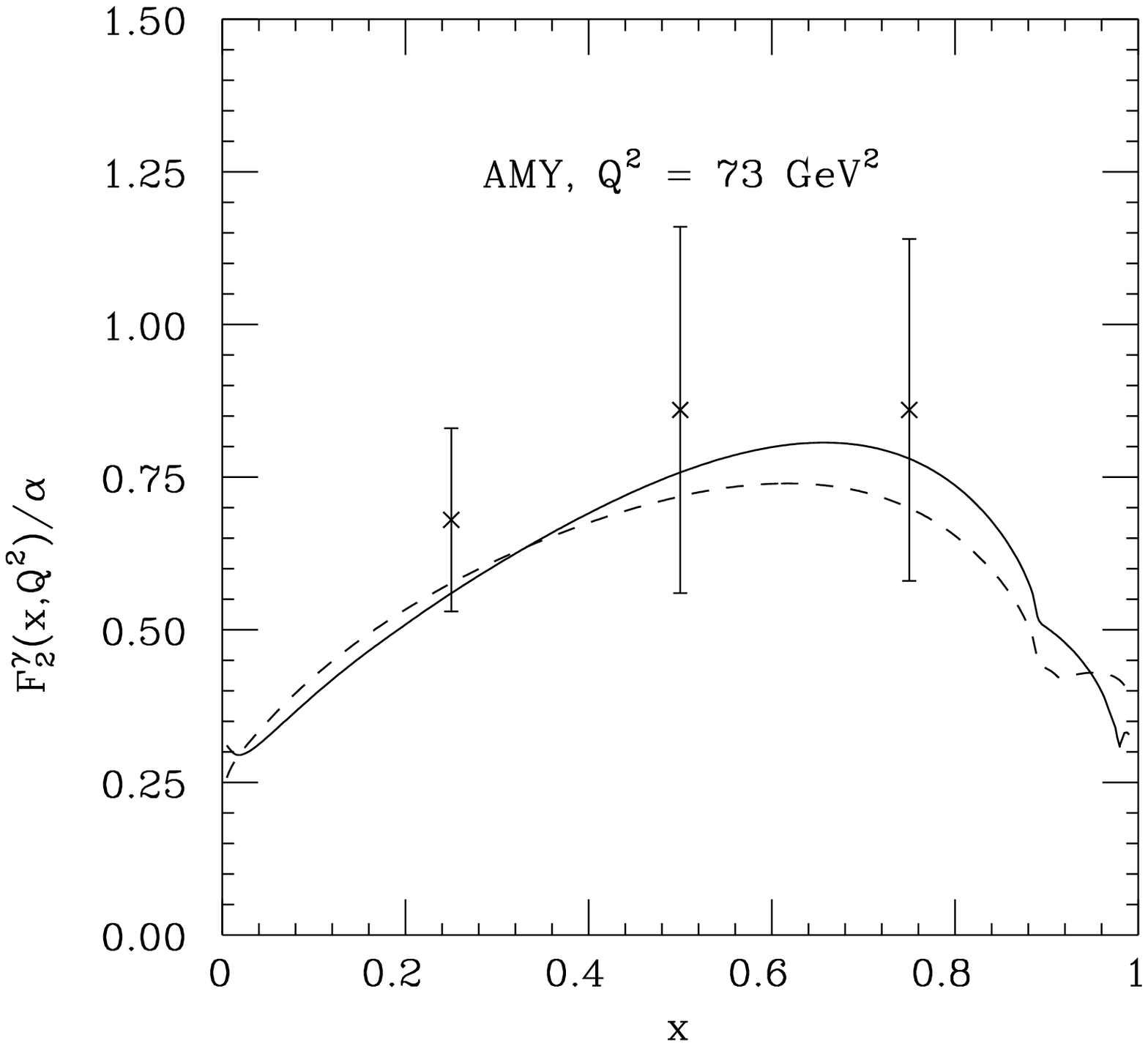}}
\end{figure}
\begin{figure}
{\hskip 0.2cm}\hbox{\epsfxsize7.5cm\epsffile{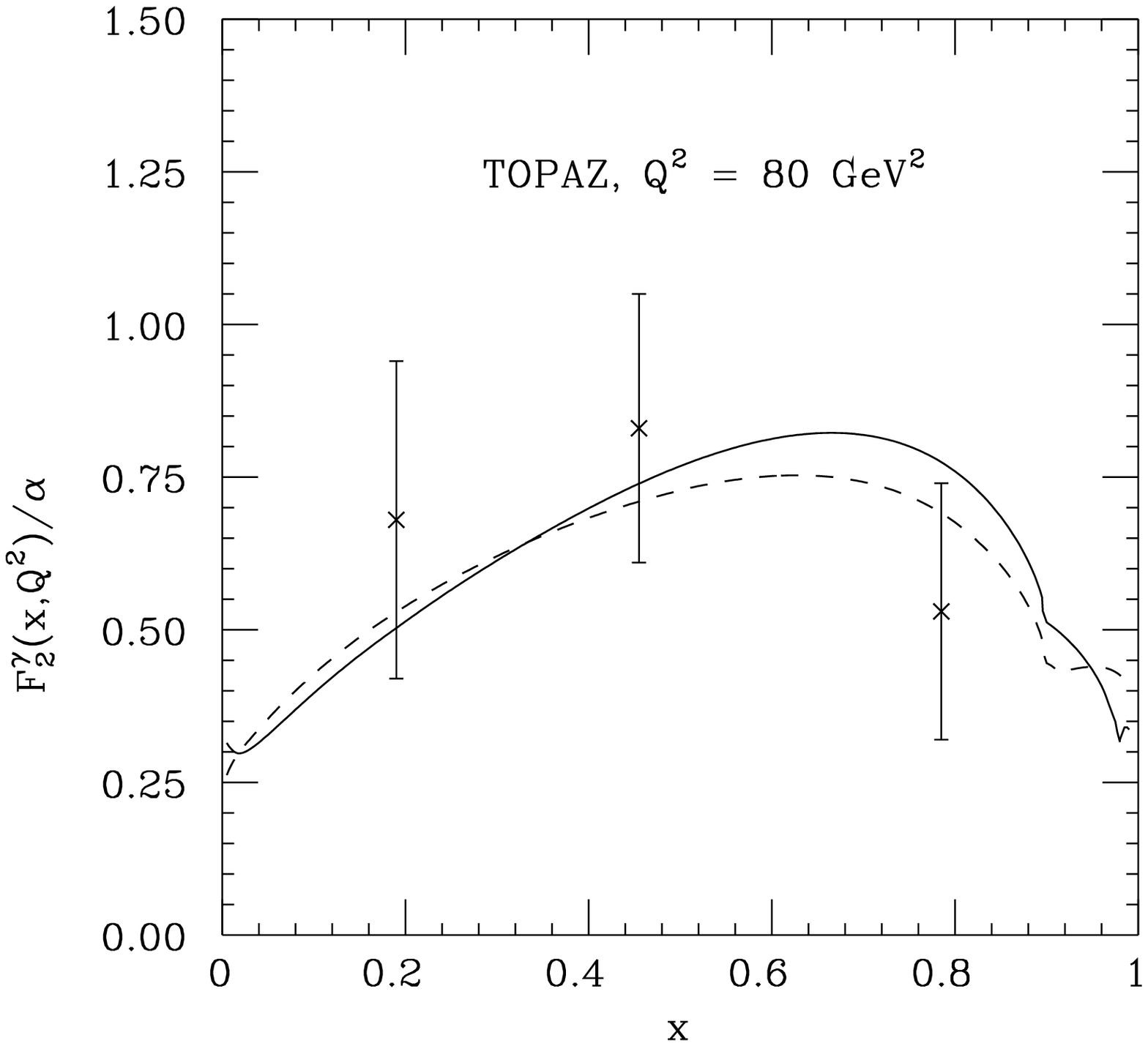}{\hskip 0.2cm}
\epsfxsize7.5cm\epsffile{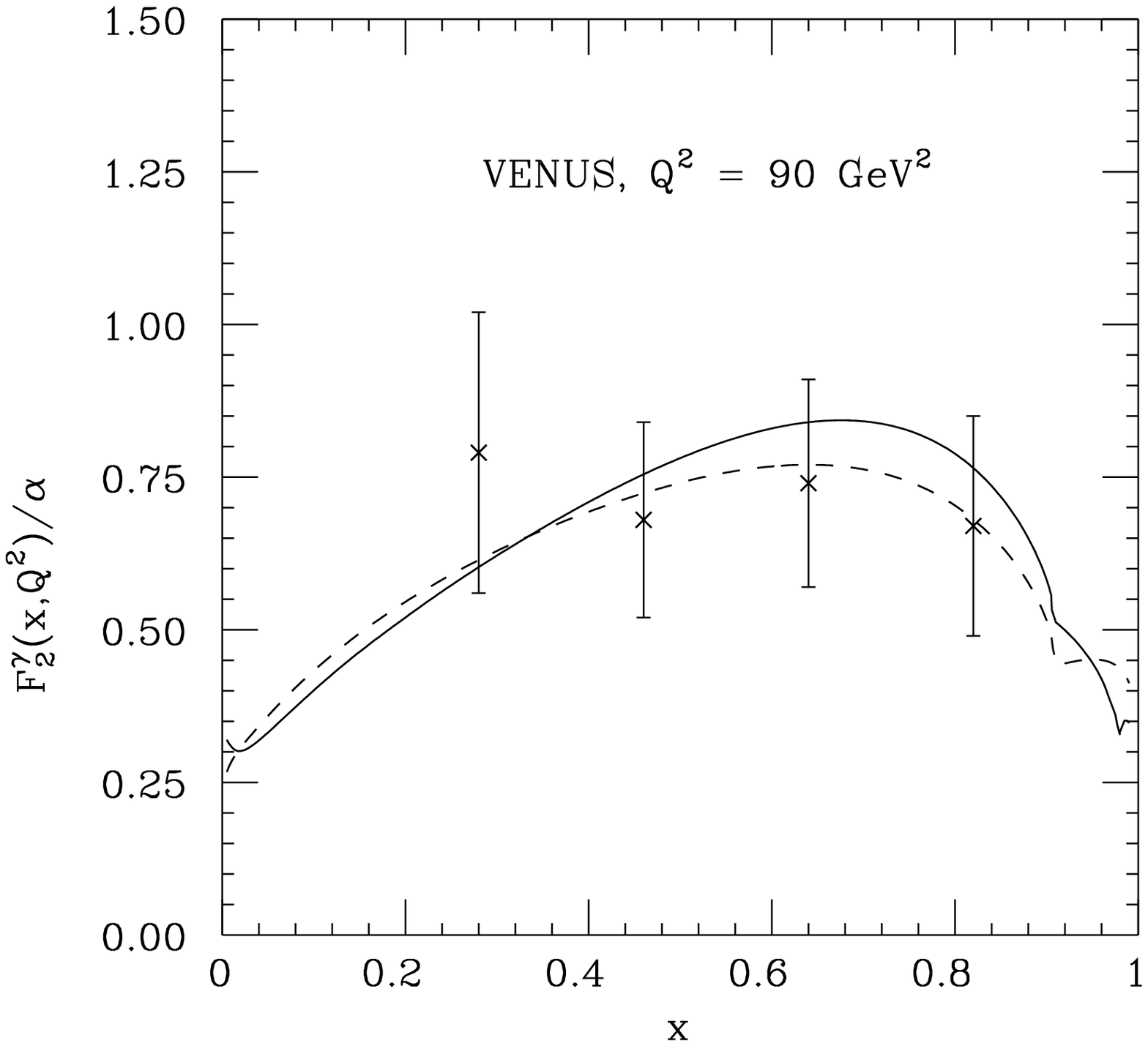}}
\end{figure}
\begin{figure}
{\hskip 0.2cm}\hbox{\epsfxsize7.5cm\epsffile{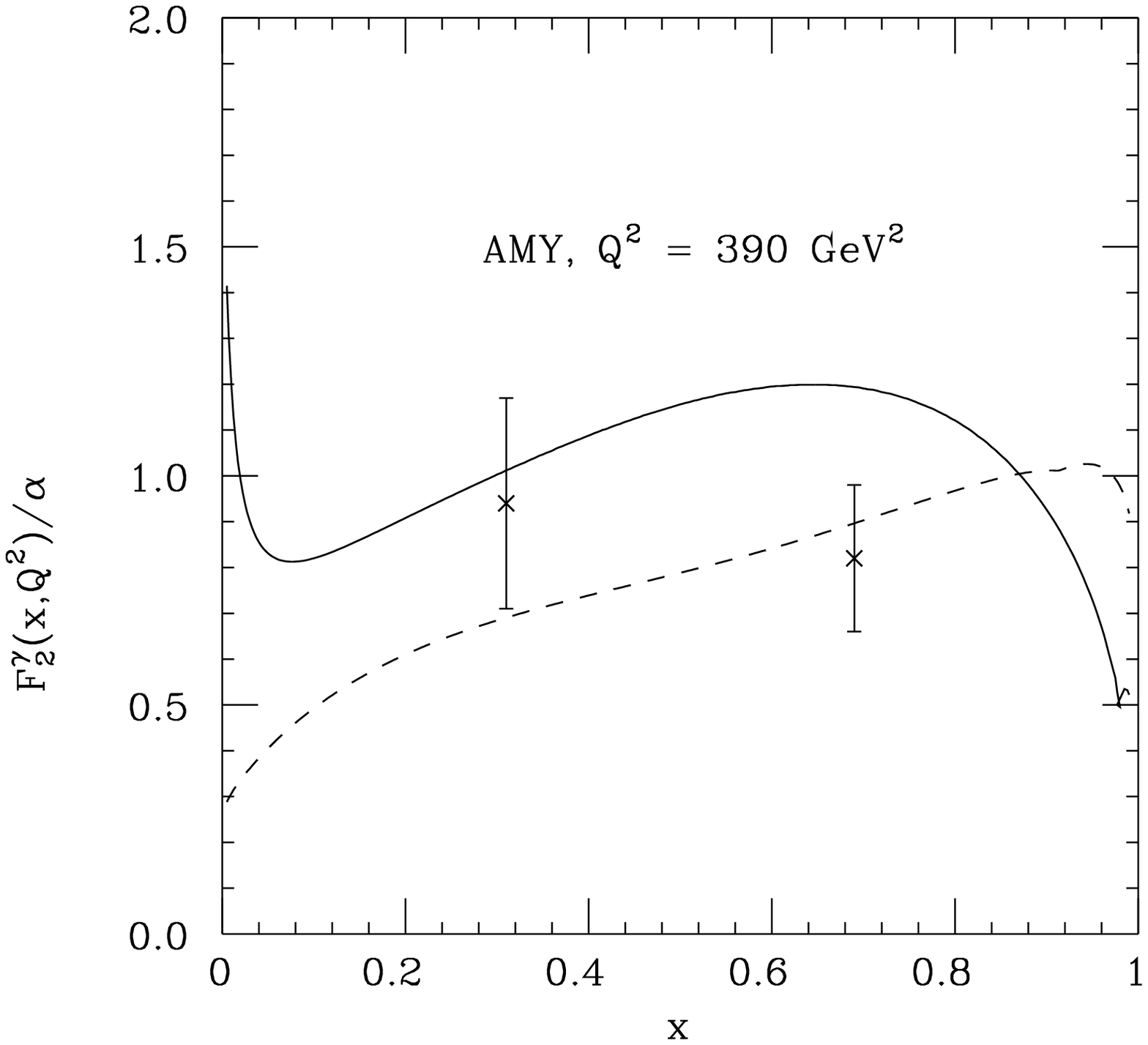}{\hskip 0.2cm}
\epsfxsize7.5cm\epsffile{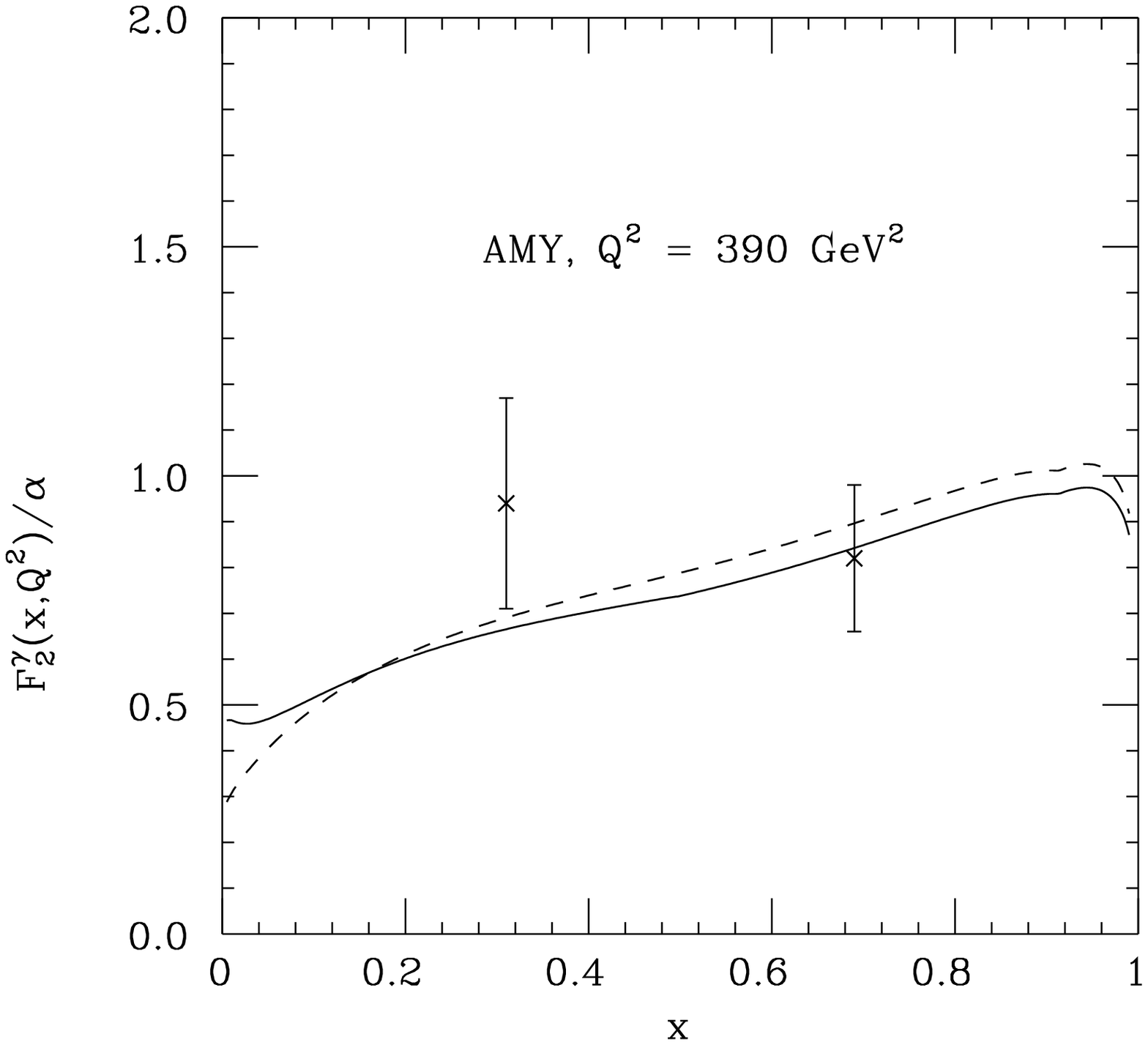}}
\fcaption{}
\end{figure}
\pagebreak
\begin{figure}
{\hskip 0.2cm}\hbox{\epsfxsize7.5cm\epsffile{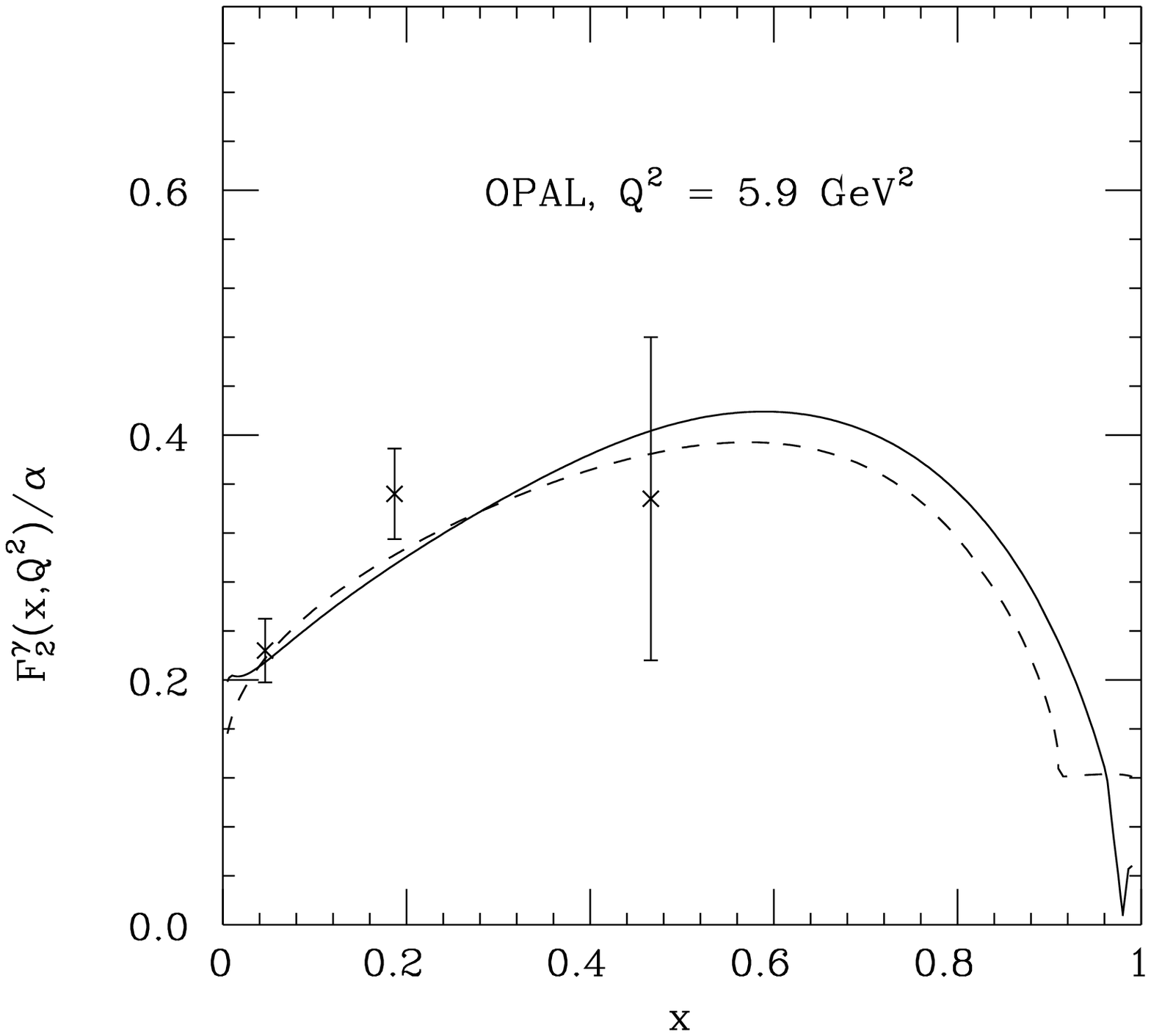}{\hskip 0.2cm}
\epsfxsize7.5cm\epsffile{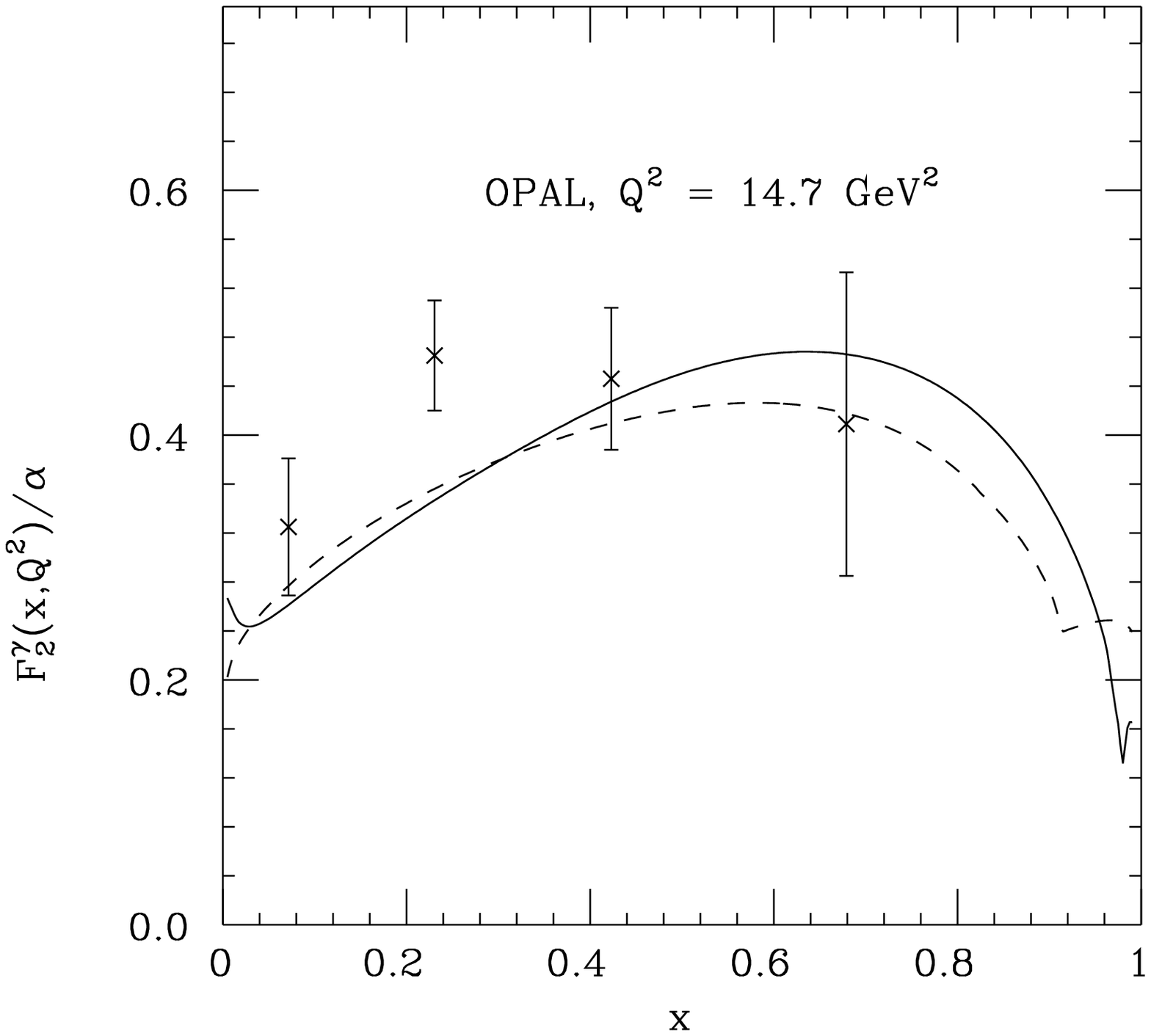}}
\end{figure}
\begin{figure}
{\hskip 0.2cm}\hbox{\epsfxsize7.5cm\epsffile{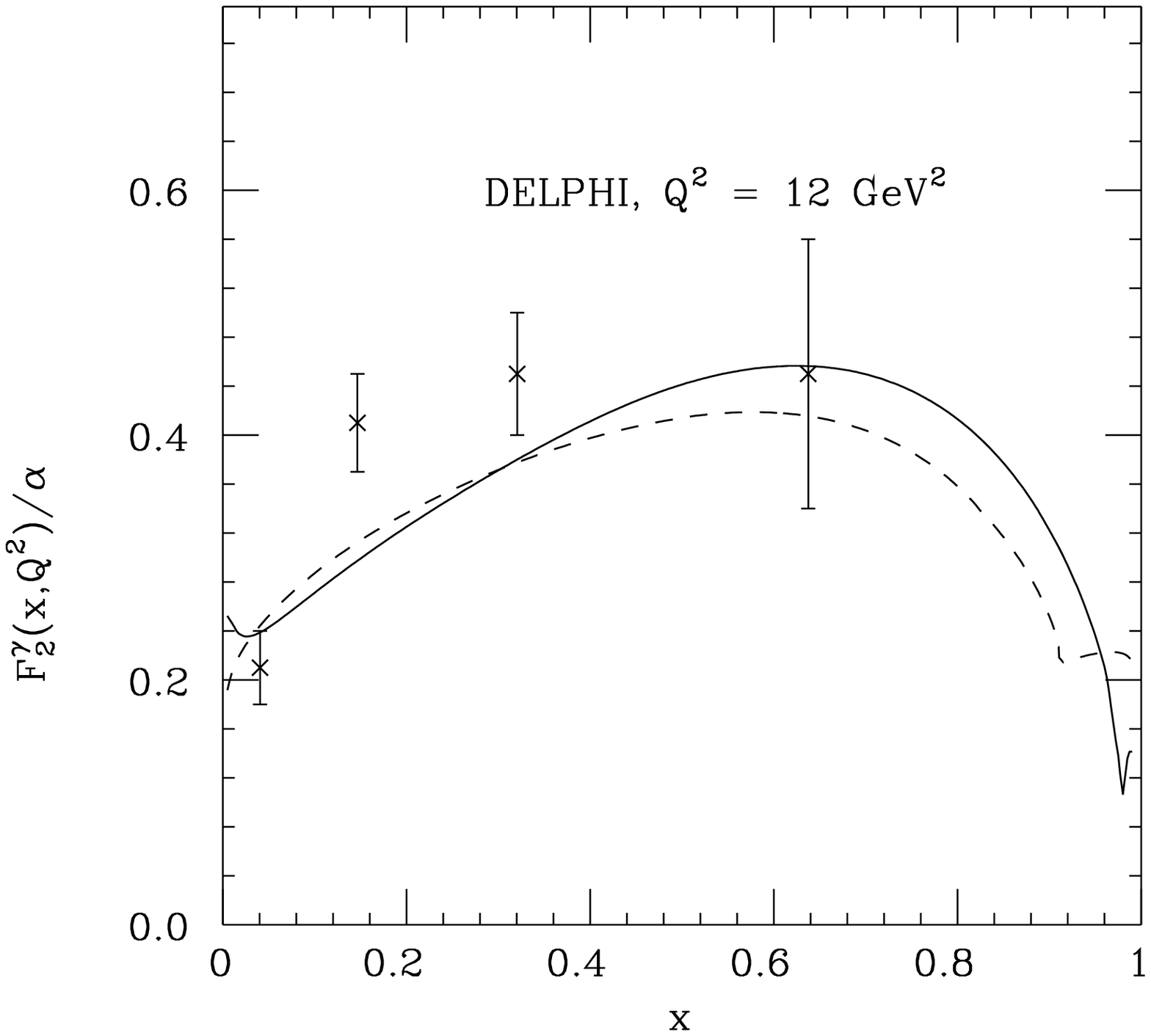}{\hskip 0.2cm}
\epsfxsize7.5cm\epsffile{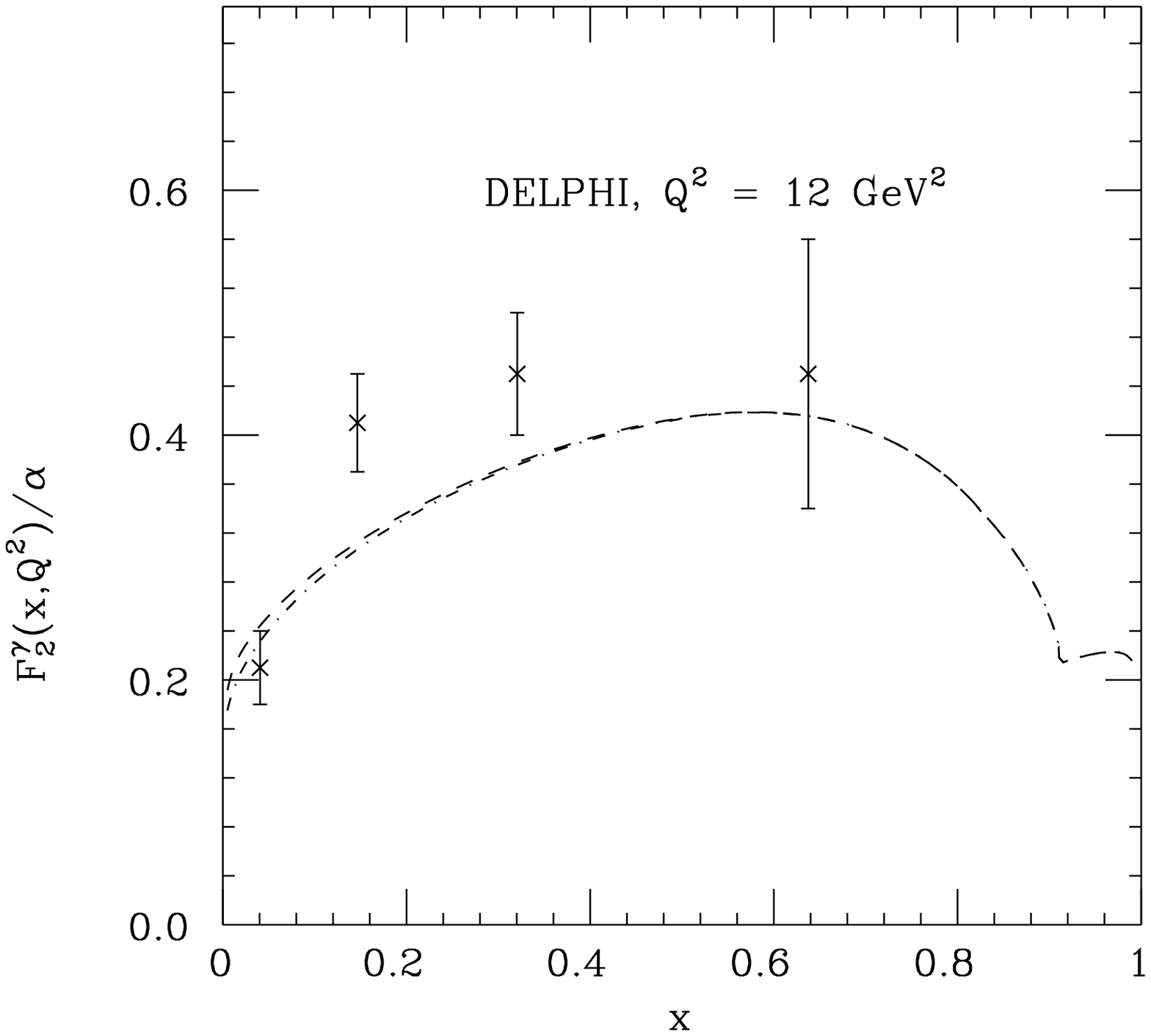}}
\fcaption{}
\end{figure}

\epsffile{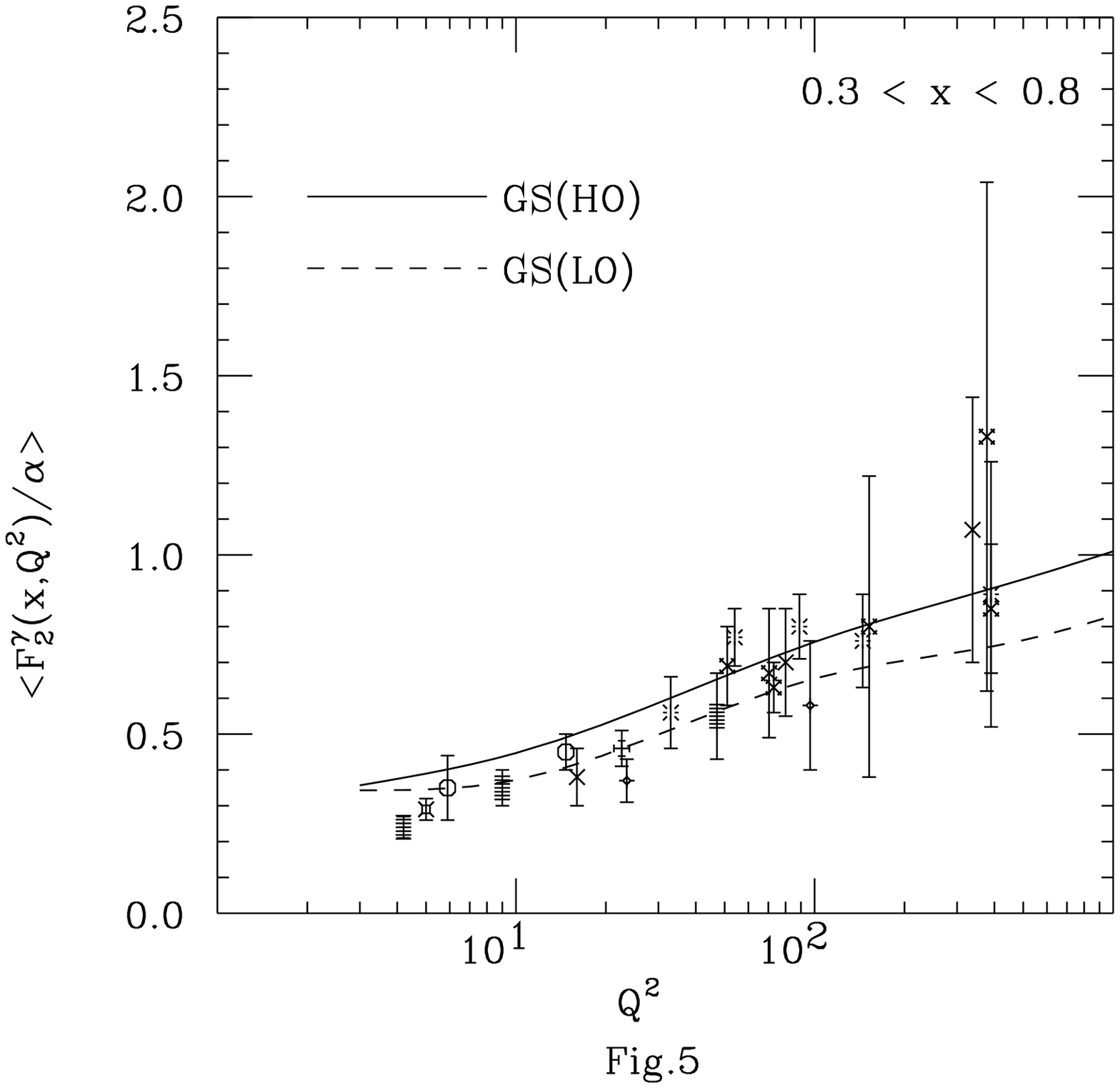}
\epsffile{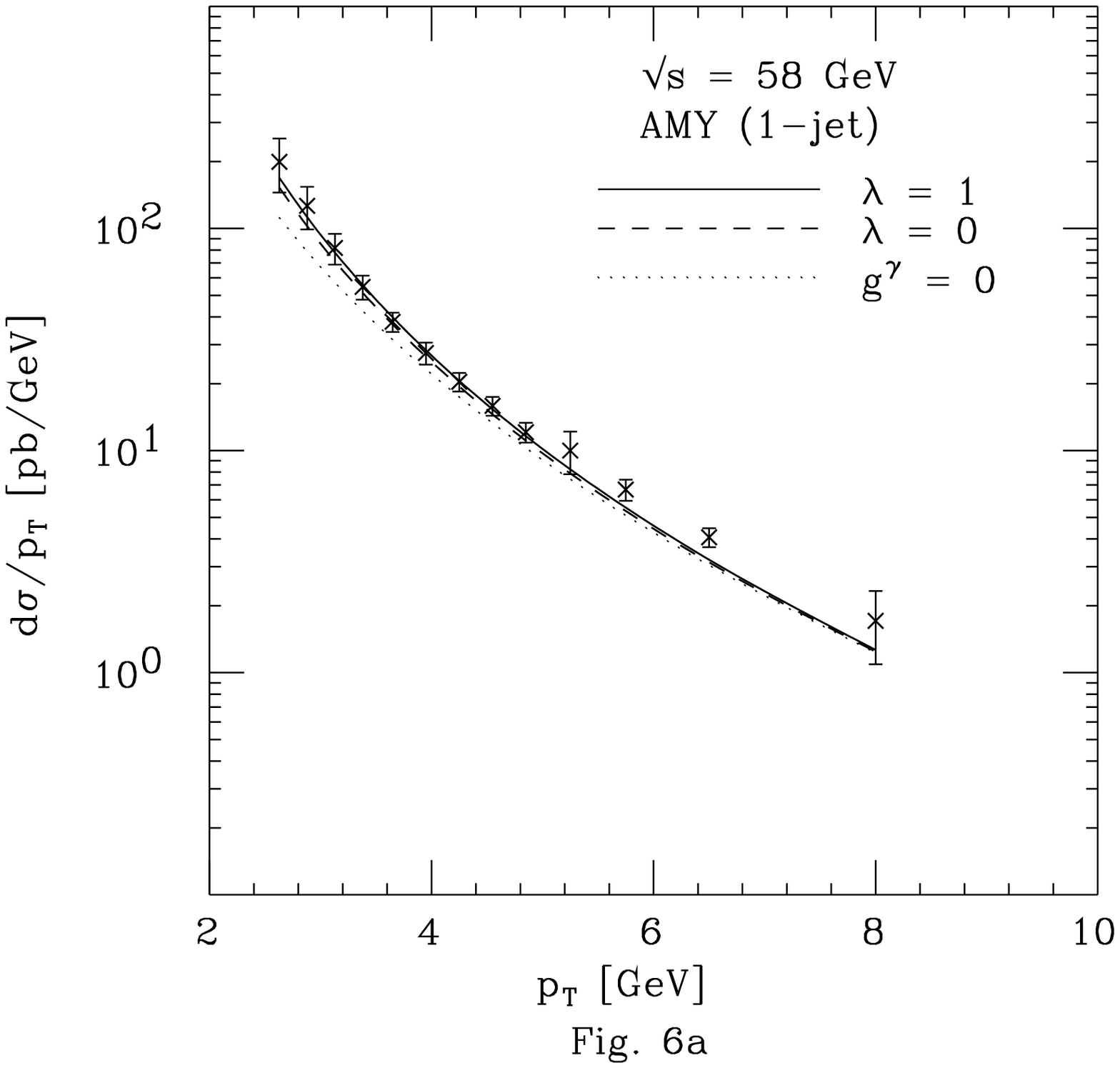}
\epsffile{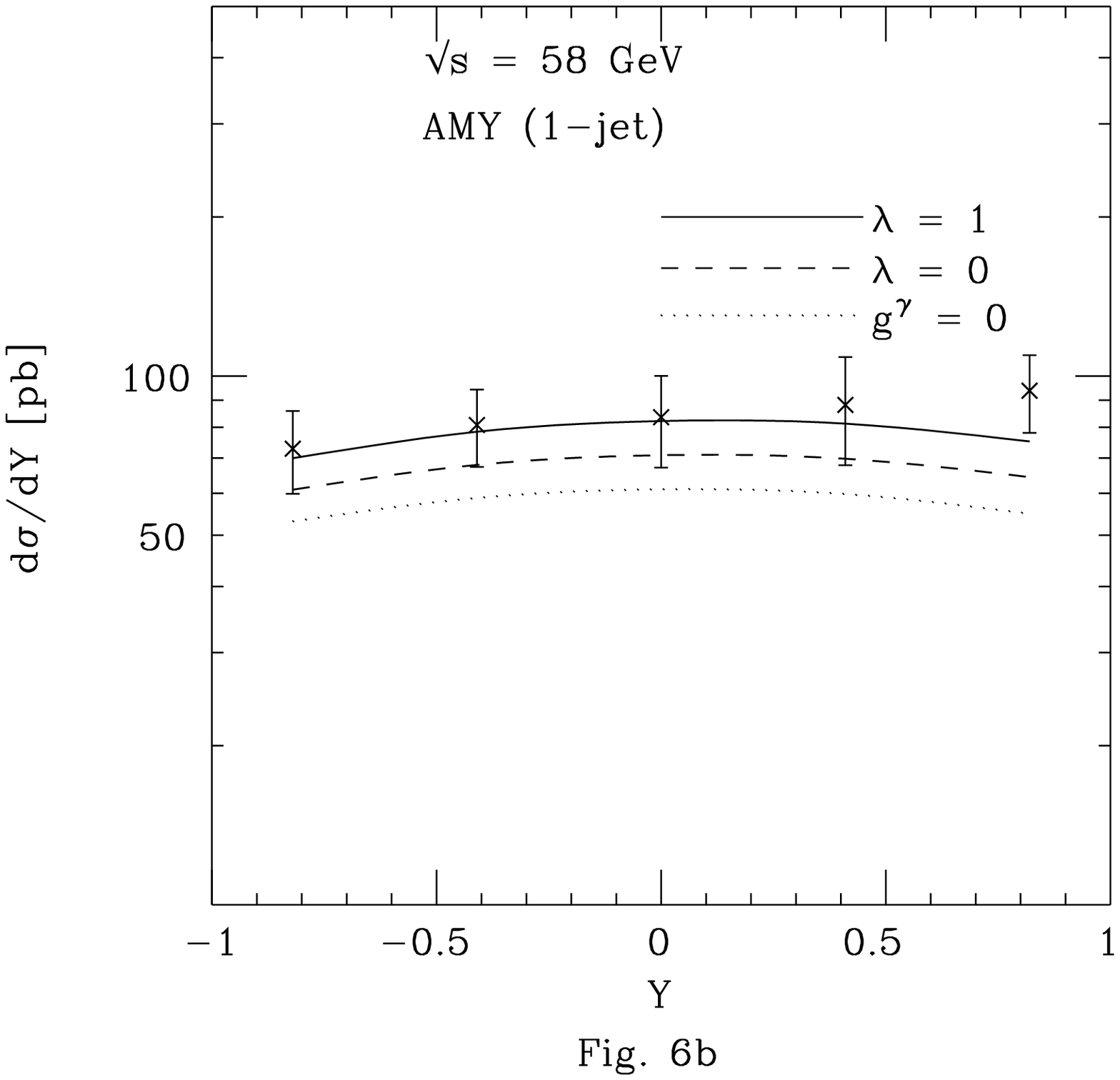}
\epsffile{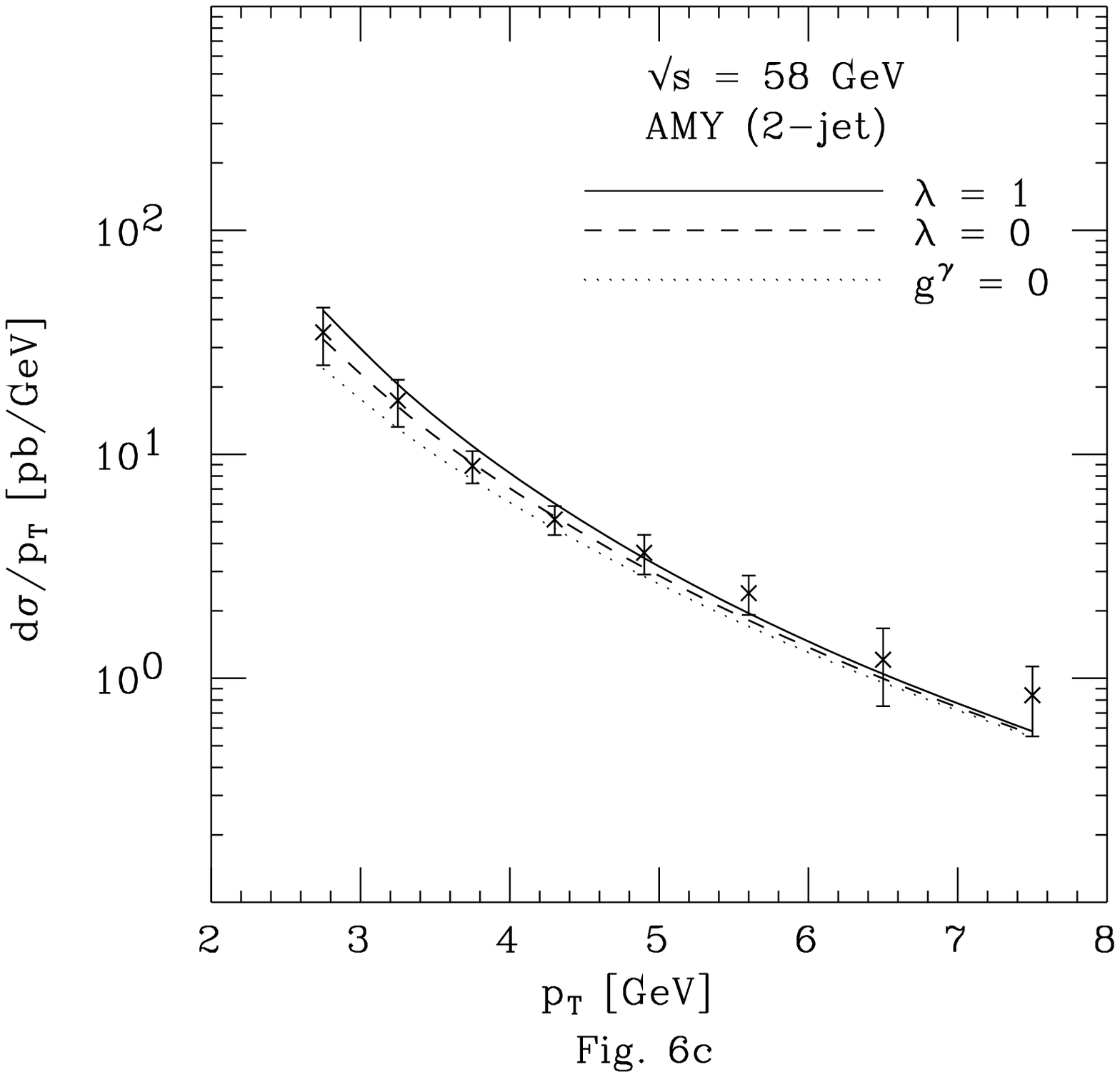}
\epsffile{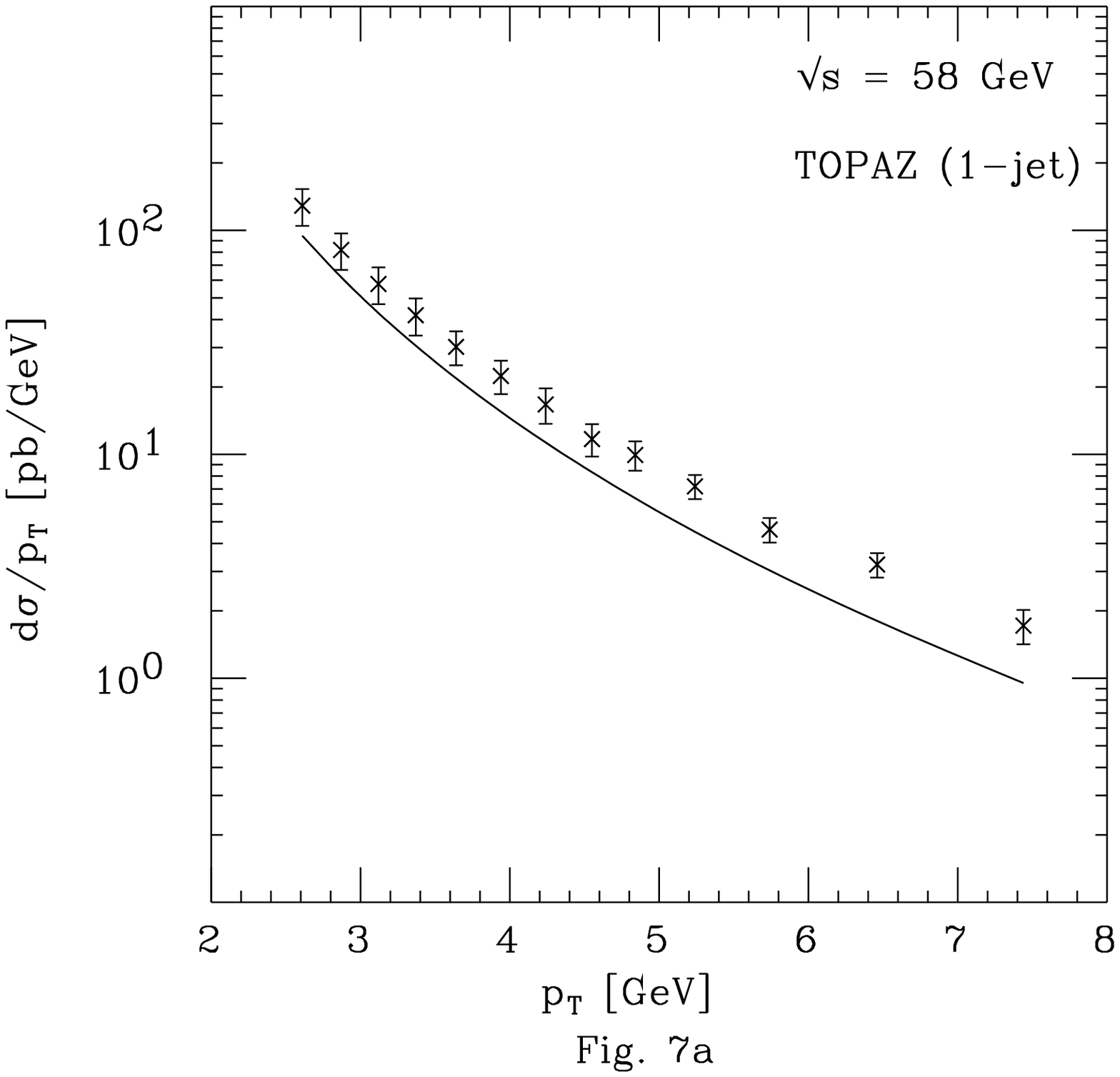}
\epsffile{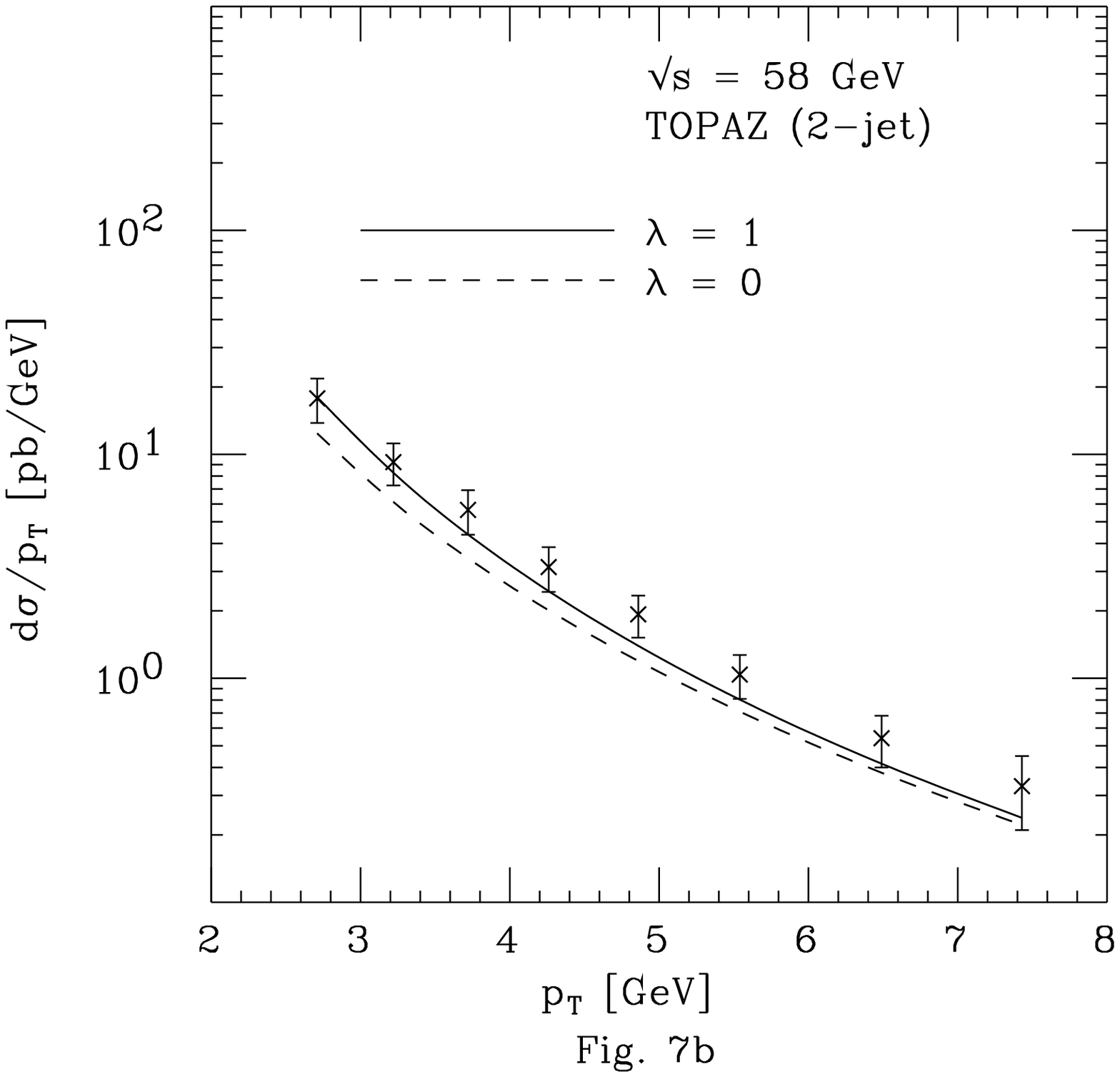}
\epsffile{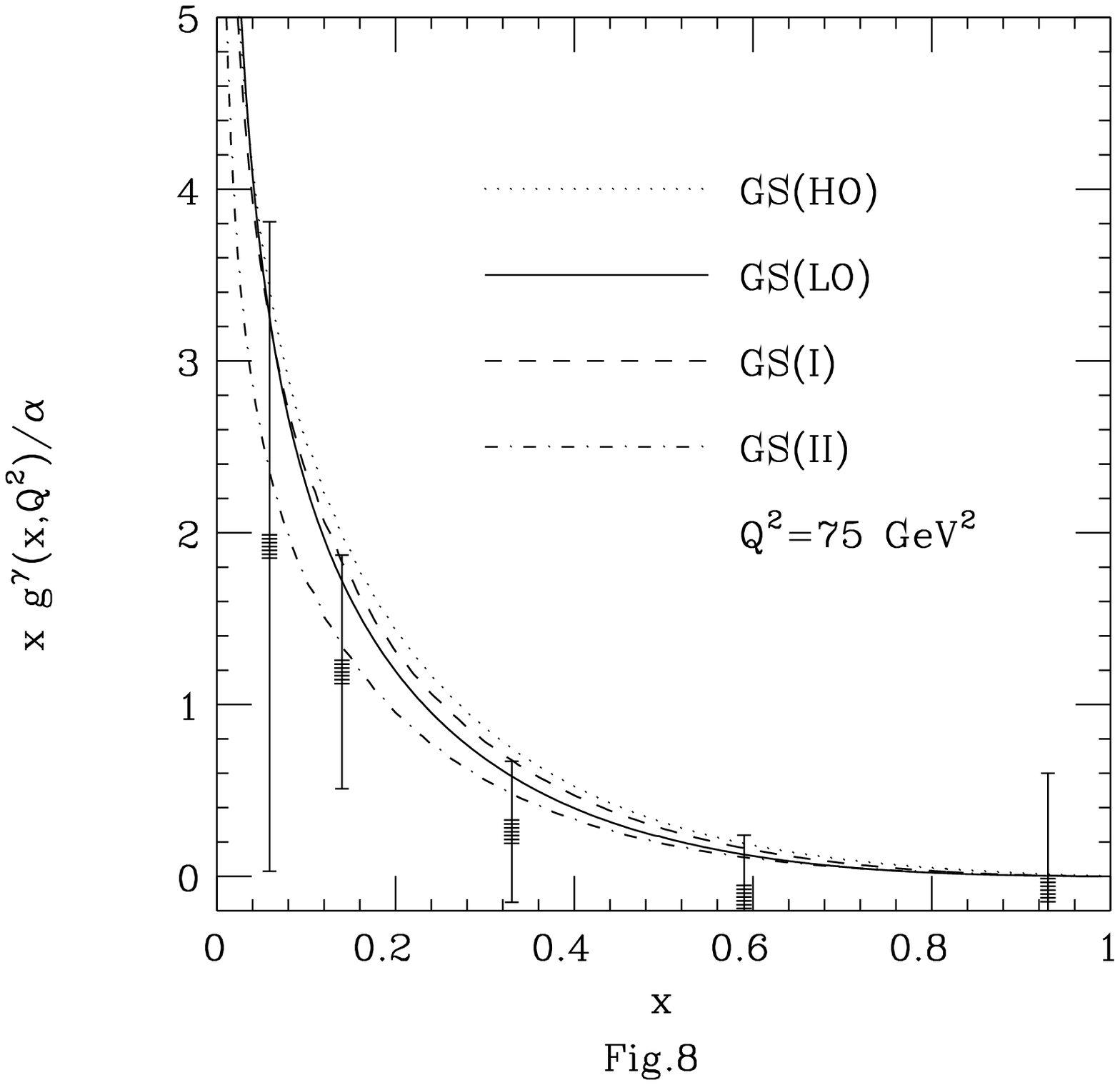}
\end{document}